%% file: ms.tex
\documentclass[sigplan,10pt]{acmart}
\usepackage{gwsys}
\usepackage{multirow}

\input{sections/macros}
\begin{document}

\date{}

\title{\huge \bf Efficient, Dynamic Multi-tenant Edge Computation in EdgeOS}

\author{Yuxin Ren}
\affiliation{\institution{The George Washington University}}
\email{ryx@gwu.edu}

\author{Vlad Nitu}
\affiliation{\institution{EPFL Lausanne}}
\email{vlad.nitu@epfl.ch}

\author{Guyue Liu}
\affiliation{\institution{The George Washington University}}
\email{guyue@gwu.edu}

\author{Gabriel Parmer}
\affiliation{\institution{The George Washington University}}
\email{gparmer@gwu.edu}

\author{Timothy Wood}
\affiliation{\institution{The George Washington University}}
\email{timwood@gwu.edu}

\author{Alain Tchana}
\affiliation{\institution{Toulouse University}}
\email{Alain.Tchana@enseeiht.fr}

\author{Riley Kennedy}
\affiliation{\institution{The George Washington University}}
\email{rskennedy@gwu.edu}

\renewcommand\footnotetextcopyrightpermission[1]{}
\pagestyle{plain}
\settopmatter{printfolios=true}
\settopmatter{printacmref=false}

\maketitle

\thispagestyle{empty}

\subsection*{Abstract}
\input{sections/abstract}

\section{Introduction}
\label{s:intro}
\input{sections/intro}

\section{Motivation}
\label{s:motiv}
\input{sections/motiv}
\section{Design}
\label{s:design}
\input{sections/design}

\section{Implementation}
\label{s:impl}
\input{sections/impl}

\section{Evaluation}
\label{s:eval}
\input{sections/eval}

\section{Related Work}
\label{s:related}
\input{sections/related}

\section{Conclusions}
\label{s:conc}
\input{sections/conc}

\bibliographystyle{ACM-Reference-Format}
\bibliography{osdi18eos,tim}



\end{document}

%% file: sections/macros.tex
\usepackage{xspace}

\newcommand{\cut}[1]{}

%% file: sections/abstract.tex

In the future, computing will be immersed in the world around us -- from augmented reality to autonomous vehicles to the Internet of Things.
Many of these smart devices will offer services that respond in real time to their physical surroundings, requiring complex processing with strict performance guarantees.
Edge clouds promise a pervasive computational infrastructure a short network hop away from end devices, but today's operating systems are a poor fit to meet the goals of scalable isolation, dense multi-tenancy, and predictable performance required by these emerging applications.
In this paper we present EdgeOS, a micro-kernel based operating system that meets these goals by blending recent advances in real-time systems and network function virtualization.
EdgeOS introduces a Featherweight Process model that offers lightweight isolation and supports extreme scalability even under high churn.
Our architecture provides efficient communication mechanisms, and low-overhead per-client isolation.
To achieve high performance networking, EdgeOS employs kernel bypass paired with the isolation properties of Featherweight Processes.
We have evaluated our EdgeOS prototype for running high scale network middleboxes using the Click software router and endpoint applications using memcached.
EdgeOS reduces startup latency by 170X compared to Linux processes and over five orders of magnitude compared to containers, while providing three orders of magnitude latency improvement when running 300 to 1000 edge-cloud memcached instances on one server.

%% file: sections/intro.tex
There is a growing desire to deploy software services closer to users.
Cellular providers must run mobility management services near customers to properly maintain connectivity for cell phone users.
The Internet of Things foretells the deployment of billions of devices producing data streams, which often require processing close to the data source to avoid excess bandwidth consumption in the network core.
Latency sensitive cyber physical systems desire communication and processing at millisecond scale, preventing the use of centralized cloud infrastructures.
These applications and many more demand an efficient and scalable ``edge cloud'' infrastructure, where computational resources are available on demand, as close to users as possible.

Unfortunately, edge clouds pose major challenges for traditional operating system and virtualization architectures.
First, an edge cloud must support dense multi-tenancy---each edge cloud site is expected to be a tiny fraction of the size of a centralized cloud, yet it may need to host many {\em carefully isolated} services for the users connected to it.
Thus rather than run thousands of servers each supporting a dozen services in virtual machines (VMs), as is common in today's centralized cloud data centers, an edge cloud site might only run a dozen servers, each supporting thousands of diverse services.
Even lightweight virtualization platforms such as Linux Containers have trouble scaling to these extremes~\cite{manco17lightvm}.

Second, the combination of limited resources and mobile users means that edge cloud workloads are likely to see extremely high churn.
Maintaining a large number of long running yet infrequently accessed services will not be efficient in such an environment, so services will instead need to be instantiated and terminated frequently on demand.
In the extreme case, this may require dynamically starting a new service {\em for each incoming user connection}.

The overarching concerns of dense multi-tenancy and high churn are compounded by the latency sensitivity and network intensive nature of many edge cloud services.
This is particularly challenging since virtualization adds overhead for I/O tasks~\cite{gupta_enforcing_2006,hu_towards_2017}.
Recent support for HW virtualization, such as SR-IOV capable NICs, reduces virtualization layer costs, but comes at the expense of scalability (e.g., only a few dozen virtual devices per port).
Thus current OS and HW virtualization techniques lack scalability and often suffer from performance unpredictability which can be a major concern for latency sensitive applications utilizing the edge.


Prior work has investigated portions of these problems in contexts such as cloud computing, network function virtualization (NFV), or real-time systems.
Lightweight virtualization techniques based on unikernels~\cite{manco17lightvm} and hypervisor optimizations~\cite{nitu_swift_2017} have been proposed to reduce boot times, but don't address providing many isolated clients high throughput.
Recent NFV platforms achieve high throughput with the use of kernel-bypass networking, but they often trade isolation for performance~\cite{han_softnic:_2015,zhang_opennetvm:_2016}.
Similarly, predictable performance is the hallmark of real-time systems, but these systems generally rely on conservative resource overprovisioning which is counter to the goals of an efficient edge cloud. 

In this paper we explore how a clean-slate OS can provide a flexible infrastructure that can securely and efficiently support a large number of isolated services, while offering strict performance guarantees.
By using a $\mu$-kernel based design, \eos\ provides a customizable architecture tuned for network intensive workloads, while providing stronger isolation and latency guarantees than existing approaches.
\eos\ uses a ``Feather Weight Process'' (\fwp) abstraction to provide fine grained isolation at low cost, with support for \fwp\ caching to assist with fast startup under high churn.

Despite its radical design, \eos\ is able to run several common applications, including middleboxes from the Click software router~\cite{kohler00click} and the memcached key-value store.
These network functions and endpoint servers can be flexibly combined to build complex services, while still providing strong isolation for both application and network data.


\eos\ makes the following contributions:
\begin{itemize}
\item A Feather Weight Process abstraction built on a $\mu$-kernel-based OS that supports orders of magnitude greater density than prior approaches.
\item Efficient mechanisms so that message data can be securely communicated through service chains.
\item \fwp\ chain caching to support microsecond speed initialization of complex services in high churn environments.
\end{itemize}

We have implemented \eos\ by extending the Composite component-based operating system~\cite{wang15speck}.
\eos\ integrates with the Data Plane Development Kit (DPDK) to provide high performance network I/O.
Our evaluation illustrates how \eos\ can offer dramatically better scale, density, and performance predictability than traditional approaches.
We execute 1000s of \fwp s per host, instantiate them 170X faster than a Linux process, and maintain a {\tt memcached} latency under 1 millisecond even when running 600 isolated instances on a single host. 
\eos\ provides performance on par with state-of-the-art NFV platforms, while offering stronger isolation and greater agility.




%% file: sections/motiv.tex
In \eos\ we consider edge clouds in the context of 5G networks, which will allow large numbers of mobile devices to connect with low latency and high bandwidth to nearby access points~\cite{taleb_multi-access_2017}.
An access point (or perhaps a nearby telco central office~\cite{peterson_cord:_2015}) can contain an edge cloud site, \ie, a tiny data center offering compute and storage capabilities to connected devices.
Edge clouds enable requests to be serviced, filtered, and transformed before they traverse the WAN, thus avoiding computation in a centralized cloud and/or reducing core bandwidth usage.
However, given the large number of edge cloud sites, each is expected to only have a small number of servers due to space, power, and cost constraints.
Since edge clouds are likely to be deployed first by telco operators, it is expected that early use cases will focus on NFV middleboxes, such as cellular mobility management, DDoS detection, etc.
Here we discuss how the scale, churn, and performance requirements of edge clouds pose major challenges to existing platforms, motivating the need for a redesign of the underlying OS primitives and communication mechanisms.





\subsection{Multi-tenancy and Churn}
\label{ss:motiv_isol}

Given the increasing number of stakeholders that can benefit from edge cloud execution, the support for multi-tenant execution is critical.
However, today's common infrastructures built on containers or virtual machines may add prohibitive cost for edge workloads.
Though past research has increased the agility of such infrastructures by optimizing the startup/shutdown costs~\cite{lagar_cavilla09snowflock,manco17lightvm,nitu_swift_2017}, the overhead of creating and deleting isolated execution environments can still be significant.

The costs of inter-tenant isolation, especially with high {\em churn} --  the rate of client arrival and exit -- can severely limit the workloads a system can handle.
A number of factors are increasing the importance of systems that can handle an increased churn at the network edge.
\begin{inparaenum}[(1)]
\item {\em serverless} computing has been made popular by platforms like Amazon Lambda and the open-source OpenWhisk, and leverages transient computations without local permanent state to increase agility and consolidation,
\item {\em middleboxes} focus on doing efficient and low-latency network computation, and benefit from per-flow isolation,
\item the number of clients accessing the infrastructure is both increasing and becoming more transient with mobile computing~\cite{manco17lightvm}, and
\item large volumes of sporadically network-connected embedded (IoT) devices are prospected to generate a majority of the world's network traffic.
\end{inparaenum}

\head{Churn and isolation overheads.}
When new clients require isolated computation in the edge cloud, namespace, memory, and CPU isolation provide the requisite separation between tenants.
Unfortunately, even relatively efficient mechanisms such as containers rely on layers of abstraction such as the Linux Virtual File System (VFS), and management of a large number of namespaces (including those for processes, network, and shared memory) that impose significant overhead.
As the churn of a system increases, the overheads of container creation are amplified.


{
\begin{center}
    \begin{tabular}{l|llll}
    Percentile & Docker    & fork()& \eos \\ \hline
    50th       & 521       & 0.26      & 0.048    \\
    90th       & 574       & 5.8       & 0.054    \\
    \end{tabular}
\end{center}
}

The table above depicts the cost in milliseconds of leveraging various isolation facilities; we measure the time to start a minimal service and then fault in 8 pages of memory to show the unpredictability of Linux's Copy on Write (full details in Section~\ref{ss:eval-boot}).
Using {\tt docker start} can take hundreds of milliseconds due to the cost of initializing namespaces and setting up Docker metadata.
Linux {\tt fork()} has a much lower cost than Docker, but it still exhibits high variance, with the 90$^{th}$ percentile being over 20 times slower than the median.
In contrast, \eos\ improves median start time by 5X compared to Linux, and has minimal variability.
As discussed in the remainder of the paper, we can improve \eos\ by another order of magnitude by maintaining a cache of services that can be started near instantaneously.
We achieve this through lightweight, yet strong isolation mechanisms, and a clean separation of the control and data paths.

\subsection{Latency and Throughput at Scale}
\label{ss:motiv_middlebox}

Lightweight isolation mechanisms such as containers facilitate running large numbers of applications (e.g., hundreds of Docker containers per server), but they cannot provide performance predictability as the scale rises.
This leads to the second key challenge in edge infrastructures: predictable performance, particularly latency, at large scale.

\head{Scaling isolation facilities.} Unfortunately, current infrastructures suffer poor performance not only under churn, but also at high scale.
Both VMs and containers see overheads due to the expense of traversing the host's software switch to determine the appropriate destination to deliver incoming data to.
This is exacerbated with new convenient, yet expensive, networking abstractions such as overlay networking provided by Docker.
While an approach such as SR-IOV can provide high performance networking to VMs or containers, it does so by dedicating virtual hardware functions that are a limited resource, preventing high scalability.

\begin{figure}[!t]
\vspace{-1em}
\begin{center}
\includegraphics[width=3in, trim={0cm 0cm 0cm 0cm}]{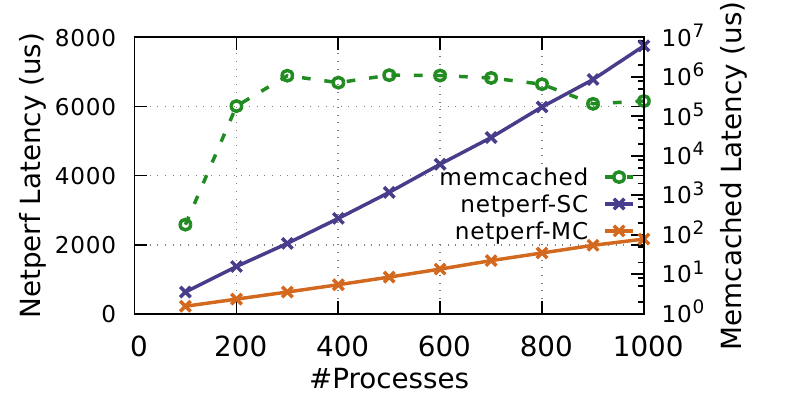}
\end{center}
\vspace{-1.2em}
\caption{\small
Round-trip latency of $N$ {\tt netperf} or {\tt memcached} instances.
Compared with the 1ms round-trip of 5G networks, {\tt netperf} latencies represent a 2x/8x latency increase using one/sixteen cores, while {\tt memcached} exhibits a 1000x latency increase.
}
\label{fig:scale_motiv}
\vspace{-2.25em}
\end{figure}

To evaluate the latency behavior of today's infrastructure, we adjust the number of {\tt netperf} servers sharing a single core ({\tt netperf-SC}) or spread across multiple cores ({\tt netperf-MC}), and the number of {\tt memcached} instances spread across multiple cores.
A second, well provisioned host transmits traffic to the test server over a 10 Gbps link.
The overhead, even in a prevalent and widespread system such as Linux, can be significant.
Using multiple cores still cannot achieve ideal latency due to poor scalability as shown in Figure~\ref{fig:scale_motiv}.
Real applications such as memcached are quickly overwhelmed and can only support a hundred or fewer instances (full details in Section~\ref{ss:eval-memcached}).
This illustrates the inability of existing OS isolation mechanisms to provide fine grained performance isolation at high scale.
\eos\ is designed to support isolation with both high scalability and predictability.

%% file: sections/design.tex
\begin{figure}[!t]
    \vspace{-1em}
    \centering
    \includegraphics[width=.95\columnwidth]{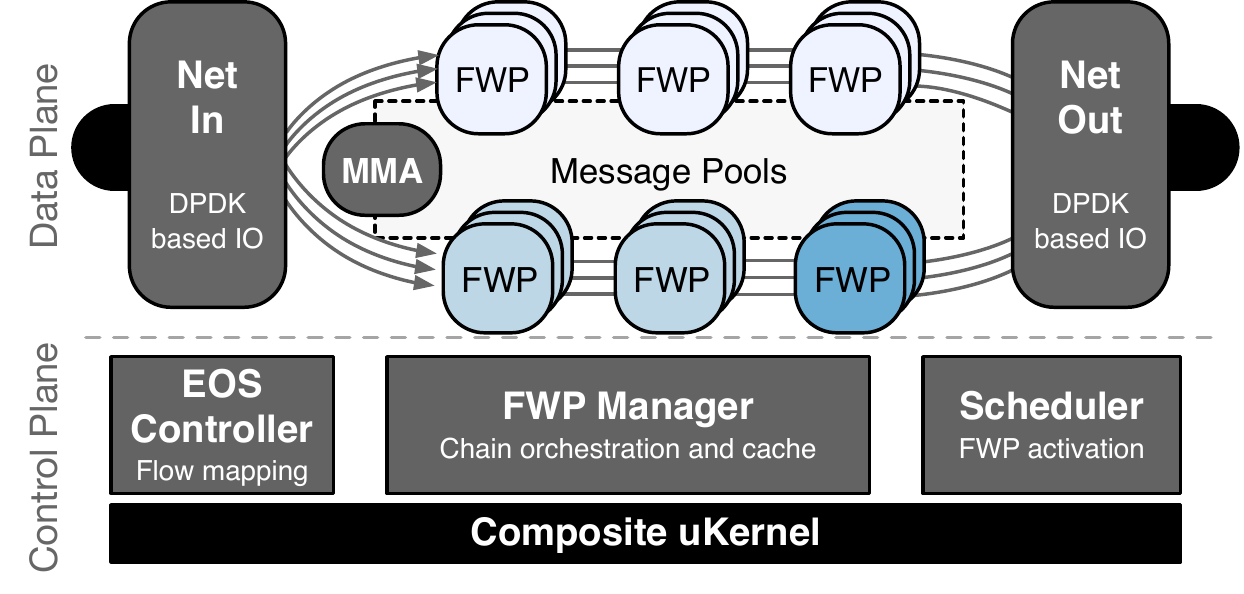}
    \vspace{-1em}
    \caption{\small \eos\ Control and Data Plane Architecture}
    \label{fig:eos-overview}
    \vspace{-1em}    
\end{figure}

As shown in Figure~\ref{fig:eos-overview}, \eos\ is designed around:
1) DPDK-based IO gateways that efficiently receive and send packets with kernel-bypass,
2) a Feather-Weight Process (\fwp) abstraction that provides fine grained isolation at low cost,
\cut{
  2) efficient communication mechanisms that can be customized to meet the security and performance requirements of each \fwp, and
}
3) a Memory Movement Accelerator (MMA) that securely copies messages between \fwp s arranged in chains, and
4) a control plane that manages the \fwp-based data plane by providing the high level policies, and offering management functions like \fwp\ template caching for fast startup.

\subsection{FWPs for Lightweight Isolation}
\label{ss:fwp}

Traditional UNIX processes maintain not only memory protection using virtual address space page-tables, but also additional abstractions including file system (FS) hierarchy visibility, file descriptor namespaces, and signal status.
Further, 
mechanisms optimized for {\tt fork} performance such as copy-on-write, and for {\tt exec} performance such as demand loading, add unnecessary {\em and unpredictable} overheads.

\begin{figure}[!t]
    \centering
    \includegraphics[width=.95\columnwidth]{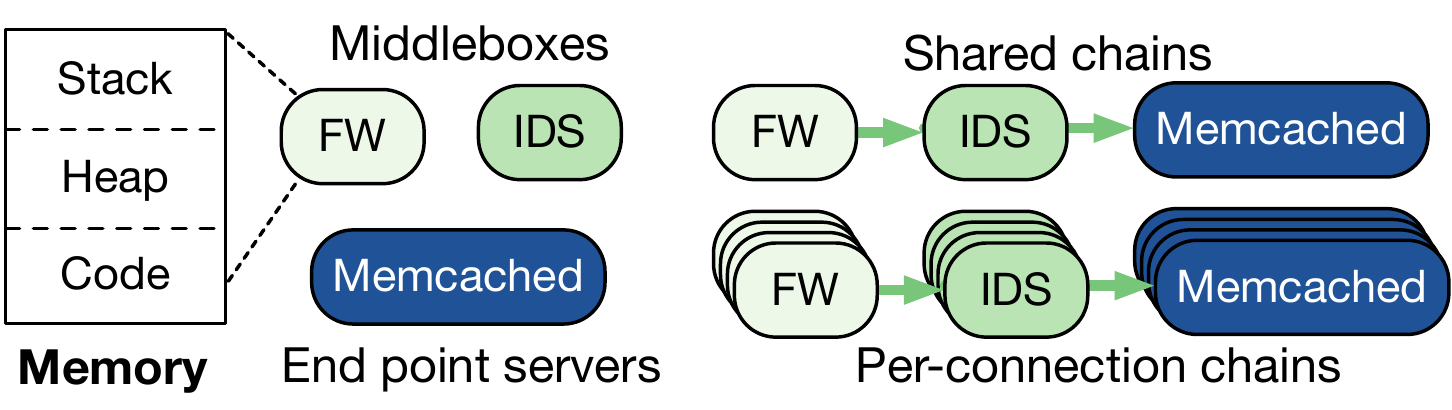}
    \vspace{-1em}
    \caption{
      {\small \fwp s can be middleboxes (\eg\ Firewalls or Intrusion Detection Systems) or endpoints (\eg\ memcached), and can be composed into chains or even replicated for every new client.}
    }
    \label{fig:fwp-cap}
    \vspace{-1.75em}
\end{figure}

In contrast, Feather-Weight Processes (\fwp s) in \eos\ are a minimal abstraction wrapping only memory and a small set of simple kernel resources.
This is partially motivated by the growing usage of stateless computation and the adoption of middlebox network functions into cloud infrastructures, signaling a growing prevalence of services that depend on external databases to store persistent state.
This enables a very tightly constrained execution environment that focuses mainly on the communication of messages (\eg\ network packets) between many, possibly untrusting \fwp s.
\eos\ optimizes around this trend.
As shown in Figure~\ref{fig:fwp-cap}, \fwp s have access only to their own memory (including stack and heap), memory for storing messages, and a number of communication end-points used to ask the \eos\ system for services.
Notably absent are default access to a file system, dynamic linking facilities, and high-level networking layers such as a TCP/IP stack.
The relative simplicity of the \fwp\ abstraction enables the efficient start-up and tear-down of computation in response to client demands.

\begin{table*}[t]
  \vspace{-1em}
  \footnotesize
  \centering
  \begin{tabular}{|l|l|}
    \hline
    Function & Description \\
    \hline\hline
    {\tt eos\_postinit(fn\_t callback, void *data)} & Provides the function that is triggered after \fwp\ initialization has completed \\
    {\tt eos\_receive\_fn(fn\_t callback, void *data)} & Sets the callback function invoked on each message reception; \\
                                                     & that function is passed both the message, and its source end-point \\
    \hline
    {\tt msg\_t eos\_recv(rcv\_ep\_t)} & A lower-level API for retrieving a message from an end-point ({\tt ep}) \\
    {\tt eos\_send(send\_ep\_t, msg\_t)} & Send a message to an egress end-point \\
    \hline
    {\tt msg\_t eos\_msg\_alloc(size\_t)} & Allocate a new message in {\em message} memory \\
    {\tt eos\_msg\_free(msg\_t)} & Free a message in {\em message} memory ({\tt eos\_send} is much more common) \\
    {\tt eos\_sbrk(size\_t)} & Allocate {\em local} memory into the heap \\
    \hline
  \end{tabular}
  \caption{\small \fwp\ Programming Interface.}
  \label{tbl:api}
  \vspace{-3em}
\end{table*}

\head{Resource access control.}
Access to all of an \fwp 's resources relies on capability-based access control~\cite{dennis66capabilities} using kernel-mediated references, removing any ambient authority~\cite{miller03cap_myths}.
These resources include the message pool that is used to receive and send data, communication end-points used to trigger the message communication, and synchronous communication end-points to request operations from system-level services.
These capabilities restrict messages to be only between \fwp s in defined {\em chains}, which can be shared by many clients, or instantiated on demand for each new connection (see Figure~\ref{fig:fwp-cap}).
\fwp s are provided minimal sets of resources consistent with the principle of least privilege~\cite{saltzer75info_protection}, which, paired with strict resource management, enables the scalable execution of isolated computations for many tenants.

\head{Programming API.}
\fwp 's primary focus is on processing of data streams (e.g., network packets), their programming API focuses around event notification and the reception and transmission of messages, summarized in Table~\ref{tbl:api}.
Each \fwp\ provides a callback at initialization that is triggered upon message reception.
Memory allocation functions distinguish between standard local memory (following a {\tt malloc}-based interface) and message memory which is integrated with the communication system.
While our current implementation uses a single thread per \fwp, the underlying \cos\ system supports hierarchical scheduling~\cite{parmer11hires}, which could be adapted for multi-threaded \fwp s.

\head{Rethinking processes for scalable isolation.} It is important to contrast the isolation properties and programming model of \fwp s with those of existing abstractions such as containers~\cite{price04zones,docker} and virtual machines~\cite{dragovic03xen}.
While containers rely on process abstractions for memory isolation, they add namespace partitioning, and resource rate consumption limitations~\cite{banga99resourceCont}.
They rely on the system call layer and Linux's monolithic kernel, thus have a large Trusted Computing Base~\cite{saltzer75info_protection} whereby a single bug in the large kernel can compromise isolation.
In contrast, virtual machine hypervisors expose an interface to virtual machines that mimics the native hardware, or is extended to include paravirtualization extensions~\cite{dragovic03xen}.
The hypervisor is often smaller and has a smaller attack surface compared to the extended POSIX interface of a system like Linux.
Virtual machines are often scheduled by the hypervisor as a collective abstraction of their applications using a virtual CPU (VCPU), thus focusing on inter-VM isolation.

In contrast to approaches that support a standard API (\eg\ POSIX or x86), the \fwp\ abstraction focuses on {\em minimizing the \fwp\ API down to the bare necessities required for network intensive edge computations}.
The API is focused on enabling different \fwp s to coordinate and compose for complex functionality -- similar in concept to UNIX pipelines.
In this way, \eos\ shares the philosophical design of $\mu$-kernels to ``a concept is tolerated inside the $\mu$-kernel only if moving it outside the kernel...would prevent the
implementation of the system's required functionality''~\cite{liedtke95l4}, but extends it to the core edge computing primitives.
\eos's system services focus on simplicity of implementation and are limited to scheduling, inter-core coordination, low-level network interfaces, \fwp s, and the capability-based access control to scope access to each.
The obvious downside of this approach is decreased legacy support.
However, we have successfully ported the Click software router and {\tt memcached} key value store to \eos.
Further, we have prototype implementations of POSIX unikernels~\cite{madhavapeddy13unikernels} (based on NetBSD {\tt rumpkernel}s
), but a discussion of these is beyond this paper's scope.

\eos ' design departs from heavyweight VM or container abstractions to enable scale and minimize the width of the system API to increase security.
Though process abstractions have often been cast aside in favor of VMs~\cite{martins14clickos}, containers~\cite{zhang_opennetvm:_2016}, or language-based techniques~\cite{panda16netbricks}, \eos\ demonstrates that simplified process abstractions with tailored minimal APIs and focused optimizations for churn and communication, can scale to a large number of tenants and clients while maintaining strong isolation for edge computation.

\subsection{Data-Plane and Communication}
\label{ss:data_plane}

Receiving and transmitting packets with the NIC has traditionally required kernel intervention to manipulate the hardware.
\eos\ embraces the recent trend towards kernel-bypass to reduce this overhead by allowing user-space management of message buffers and network card DMA rings.
Though \fwp s have isolated local memory, the memory used for message passing between \fwp s exposes a trade-off between performance and isolation.
Existing high throughput systems often eschew isolation and use shared memory to pass data by reference.
This is the design chosen by high-throughput networking stacks and software middleboxes~\cite{zhang_opennetvm:_2016,palkar15e2,panda16netbricks}.
In \eos , we leverage {\em data copying} between separate \fwp s to maintain strong mutual isolation.
Data copying can be a very expensive operation as it can dirty caches
Thus, \eos\ pairs strong isolation, with Memory Movement Accelerators (\mma s) that decouple copying from the \fwp\ fast-path.

\head{Network Gateways.}
\eos's microkernel design is a natural fit for user-space packet processing frameworks such as DPDK.
In and Out gateway services run on dedicated cores and pull packets into message pools with no kernel interactions.
Input packet processing maintains rules dictated by the control plane to match packets to a destination \fwp\ service.
Depending on the rule specification, it may be necessary to instantiate a new \fwp\ chain in order to handle the incoming request.
\fwp\ chains are a core abstraction in \eos\ as the entire chain can be created to service a new client.

\head{\fwp\ Memory and Isolation.}
An \fwp 's memory is separated into {\em message memory} that is used for message passing between \fwp s, and {\em local memory} that backs each \fwp 's data-structures.
This separation enables memory allocations to be optimized for the purpose and use of the memory.
Though future optimizations might relax isolation, \eos\ focuses on strong protection between \fwp s, and employs copying to safely transfer data between each other.
When messages are passed between \fwp s, a trusted system component must be involved as neither \fwp\ has the access rights to copy into, or from, the other \fwp 's memory.
\head{Efficient message passing with the MMA.}
A key \eos\ design is to move message copying off the fast-path of \fwp\ message processing, as we have observed that even a single in-line copy can prevent line-rate processing in many cases.
Toward this, \eos\ employs a Memory Movement Accelerator (\mma) whose focus is on efficiently copying messages between \fwp s.
The \mma\ retrieves messages from a upstream \fwp 's ring buffers, copies them and adds them into a downstream \fwp 's ring buffers, and alerts the scheduler that the destination \fwp\ needs to be activated to receive it.
The \mma\ acts as a software DMA engine to move message data between \fwp s, and runs on one or more {\em dedicated cores} in order to perform out-of-band data movement.
In contrast to long-standing networking subsystem guidance that dictates that zero-copy is necessary~\cite{eicken95unet,han2012megapipe} -- often at the price of isolation, \eos\ optimizes the \mma\ and treats it as a specialized processor that can push data significantly faster than line-rate, while maintaining strong isolation.



\subsection{Control Plane and FWP Lifecycle}
\label{ss:control_plane}

Similar to the approach taken in Software Defined Networks (SDN) and split-OS designs such as Arrakis~\cite{peter15arrakis}, \eos\ separates the data plane processing (implemented with \fwp s, \mma s, and network gateways) from control functions that determine request routing, security policies, and resource management (implemented as user space components extending the \cos\ $\mu$kernel).
As shown in Figure~\ref{fig:eos-overview}, \eos 's control plane is composed of three major components: \begin{inparaenum}[(1)]
  \item the EOS Controller that maps incoming flows to \fwp\ chains,
  \item the \fwp\ Manager that controls the lifecycle of \fwp s and optimizes their startup, and
  \item the Scheduler that determines which \fwp\ to run on each core and activates them in response to incoming messages.
\end{inparaenum}

\head{Flow matching with the \eos\ Controller.} When new requests arrive from connected client devices, they need to be routed to the appropriate \fwp\ chain.
The \eos\ Controller allows administrators to define \fwp\ chains and the packet filtering rules that specify what traffic should be routed to them.
These rules are pushed to the Net-In data plane component.
Net-In applies rules similar to SDN match-action rules: packets are split into flows based on the header n-tuple (\eg\ src/dest IP and protocol) and a rule is found that matches the flow.
The rules indicate the \fwp\ chain that will process that flow.\footnote{Our implementation currently assumes flow rules are statically preconfigured, but this could be extended to support on-demand flow lookups similar to SDN controllers, with a northbound interface to application logic that would assign a rule dynamically to each flow.}
Since our focus is on fine-grained isolation and high scale, a rule can indicate whether all flows that match the rule should be handled by a single chain, or if each flow should be given a dynamically started instance of the chain.

\head{FWP Lifecycle and Caching.}
The creation of \fwp s on the fly in response to the arrival of a new flow requires a cascade of activity: the instantiation of a set of new \fwp s (including memory initialization, kernel data-structure management, and thread creation), connecting the \fwp s together with communication channels (ring buffers, kernel end-points, and \mma\ integration), and finally, the creation of the message memory regions for the \fwp s.

\begin{figure}[t]
    \vspace{-1em}
    \centering
    \includegraphics[width=1\columnwidth]{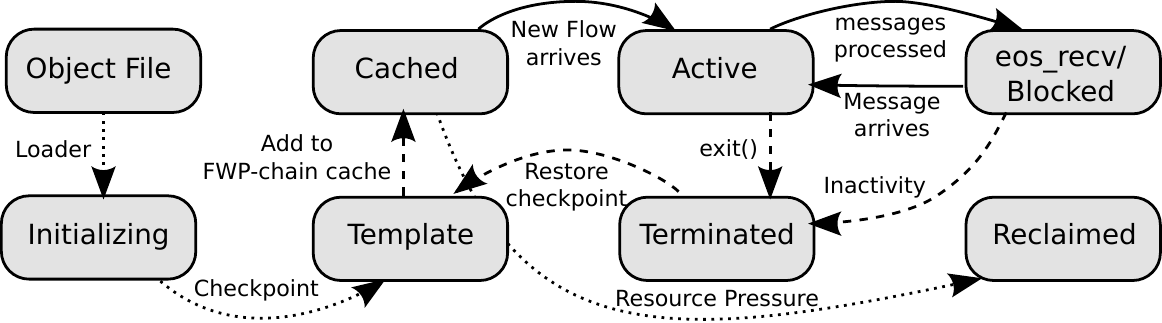}
    \caption{\small Lifecycle of a \fwp-chain:
      Dotted lines indicate \fwp\ manager operations conducted once to load and then checkpoint a \fwp-chain, or to reclaim the \fwp 's resources when memory pressure exists.
      Dashed lines indicate operations to re-initialize terminated \fwp-chains for future use.
      Solid lines are {\em data-path} operations performed by on the critical path of \fwp\ execution.
      \label{fig:lifecycle}
    }
    \vspace{-1.5em}
\end{figure}

The \fwp\ Manager orchestrates the lifecycle of \fwp\ chains, which is illustrated in Figure~\ref{fig:lifecycle}.
Similar to a Linux process, an \fwp\ starts as an object file, which must be loaded into memory.
Once execution begins, \fwp s typically perform some initialization routines (\eg, parsing configuration files and allocating initial data structures).
Rather than repeat such computation every time a new \fwp\ of the same type must be instantiated, \eos\ optimizes startup with an \fwp\ checkpoint cache.
Thus, we utilize the {\tt eos\_postinit()} API to allow \fwp s to first initialize, then to take a checkpoint that defines the state of an \fwp\ ready to process new data.

Since we anticipate many complex services will require multiple \fwp s arranged in a chain, the Manager employs a {\em \fwp-chain cache} that caches entire chains of \fwp s, their interconnections, and their message memory.
As new flows arrive, they are paired with corresponding \fwp-chains from the cache.
The selected \fwp s will be Activated, allowing them to process messages or transition to the Blocked state, before eventually Terminating when they are no longer needed.

When a \fwp\ chain terminates, the Manager reuses the chain by Restoring it back into the \fwp-chain cache.
In doing so, \eos\ must guarantee that the memory of the cached computation represents the checkpointed, post-initialization state.
As this places data-structures into a known and safe state, it ensures the integrity of future \fwp-chain instances.
\eos\ {\em avoids control operations in the data-path}, thus the Manager's checkpoint and restore operations run in parallel to \fwp\ message passing.
If memory pressure exists in the system, cached \fwp\ templates and chains are Reclaimed.


\head{Scheduling and inter-\fwp\ coordination.}
Once a set of \fwp s are activated, they are distributed across cores, and partitioned scheduling (\ie\ without task migrations) multiplexes the core's processing time.
Each scheduler requires global context on
which \fwp s are assigned on its core, and
which are runnable, 
and which are blocked awaiting messages.

Traditional systems often use direct coordination between cores via shared data-structures and explicit notification using Inter-Processor Interrupts (IPIs).
For example, Linux provides notifications to activate threads (via {\tt futex}es, or pipes) by accessing that thread's data-structure directly to see if it is already awake, and if not, an IPI is sent.
The resulting cache-coherency traffic for access to shared data-structures, then the IPI overheads, can be significant, especially if used for message notifications arriving over a network at line rate.
\fwp-chains can be spread across cores, only increasing the cost.
Motivated by these overheads, NFV platforms based on DPDK such as OpenNetVM~\cite{zhang_opennetvm:_2016} use active polling for communication between threads on different cores, thus entirely avoiding blocking.
However, as the number of processes (``network functions'' in OpenNetVM) grows beyond the number of cores, spin-based event notification is inefficient.

To avoid the large overheads of shared resources, all inter-scheduler coordination in \eos\ is via message passing.
When a \fwp-chain is activated, a message as such is sent to the scheduler controlling the core hosting the \fwp.
Additionally, when a message is sent to a \fwp, and its ring buffer is empty, a message is sent to the corresponding scheduler.
On the other hand, when an \fwp\ has processed all of its pending messages, instead of spinning awaiting more, the {\tt eos\_recv} operation will invoke the scheduler (which uses IPC to the scheduler component) asking to block.

\begin{figure}[t]
    \vspace{-1em}
    \centering
    \includegraphics[width=.95\columnwidth]{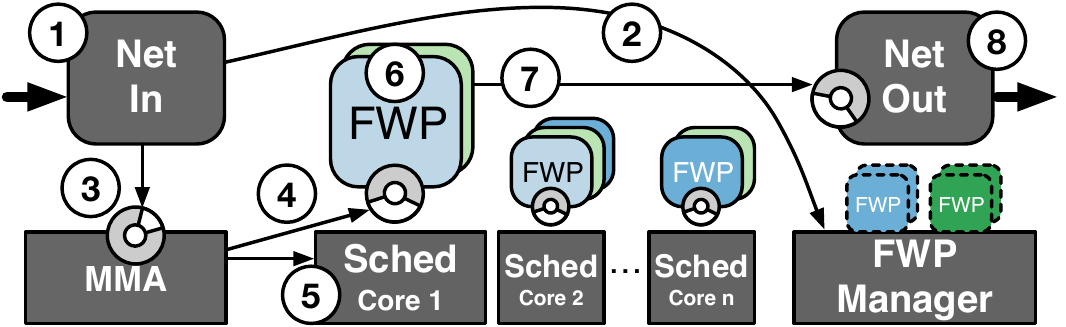}
    \caption{
      {\small \eos\ Timeline}
    }
    \label{fig:eos-timeline}
    \vspace{-1em}
\end{figure}

\vspace{1mm}
\head{\eos Timeline Summary.} Figure~\ref{fig:eos-timeline} shows the complete timeline for receiving and processing a packet.
1) A packet reception at the Net-In gateway causes a flow lookup to decide which \fwp\ chain should process the packet. 
2) If there is a miss and no \fwp\ is currently allocated, the \fwp\ Manager spawns one from its cache. 
3) A message is sent to the MMA causing it to copy the packet into the destination \fwp's pool. 
4) A message is added to the \fwp's ring and 5) the MMA messages the scheduler on the \fwp's core to activate it. 
6) The \fwp\ processes the packet and 7) asks the output gateway to DMA the packet out the NIC.

%% file: sections/impl.tex
In this section we describe how our \eos\ design is implemented and the key optimizations we make to achieve predictable, high performance.
We plan to release our source code and experiment templates for repeatable research.

\subsection{EdgeOS Implementation in Composite} 
\label{ss:cos}

\cos\footnote{\url{composite.seas.gwu.edu}} is an open source $\mu$-kernel that externalizes traditionally core kernel features into user-level {\em components} that define the resource management and isolation policies~\cite{wang15speck}.
Components interact through highly-optimized Inter-Process Communication (IPC) to leverage system logic and resources.
Similar to Eros~\cite{shapiro99eros} and seL4~\cite{elphinstone13l4}, \cos\ is based on a capability-based protection model that controls component access to kernel resources.
These resources include threads, communication end-points (synchronous and asynchronous), page-tables, capability-tables, temporal capabilities~\cite{gadepalli17tcaps}, and memory frames.
The kernel includes no scheduling policies, instead implementing schedulers at user-level~\cite{parmer08composite_sched}.
The \cos\ kernel scales up to multiple cores well as it has no locks and is designed entirely around store-free common-paths, wait-free data-structures, and quiescence for data-structure consistency~\cite{wang15speck}.

\eos\ builds on these underlying facilities to provide:
\begin{inparaenum}[(1)]
 \item \fwp\ management and caching capabilities,
 \item a DPDK-compatible userspace networking module,
 \item new communication mechanisms built around the \mma, and
 \item a scheduler that is integrated with the communication and DPDK modules.
\end{inparaenum}
\eos\ is implemented as a component consisting of these main system modules.
Co-location of these in a component is convenient and simplifies their communication, but is not necessary.
Together, they provide the abstractions to execute \fwp s as isolated components with only a limited number of synchronous communication channels to \eos\ corresponding to the functions in Table~\ref{tbl:api}.
Thus, the attack surface of any given \fwp is restricted and small.

The current \cos\ implementation is for 32 bit x86.
Though this limits the scale of the system due to memory limitations, our prototype demonstrates the core functionality of \eos.
Ports to other platforms such as ARM and x86-64 are in progress by the \cos\ developers, and we expect \eos\ would exhibit similar behavior on them.

\subsection{Feather-Weight Process Management}
\label{ss:fwpimpl}

\head{Optimized \fwp\ checkpointing.}
\eos\ caches the images of {\em chains} of \fwp\ binaries so they are ready for prompt activation.
These ready-to-execute images are {\em asynchronously} prepared, thus moving the overhead for \fwp\ preparation off the fast-path.
The cache contains full \fwp\ chains so that complete services can be quickly deployed.
The cached \fwp s represent their execution immediately following the  {\tt eos\_postinit} function, thus capturing the {\em initialized} state of a ready-to-execute \fwp.
This avoids redundant initialization computation.
For example, our Click network functions trigger the checkpoint only after loading and parsing their configuration file from disk.

However, the mechanisms to prepare \fwp-chains (in the \fwp\ Manager) still must be efficient to maintain a high churn rate.
Thus, we utilize a few optimizations:
\begin{inparaenum}[(1)]
\item the post-initialization checkpoint of the \fwp-chain is laid out contiguously in memory so that re-initializing a chain is bounded mainly by {\tt memcpy} and {\tt memset} overheads (for which we use the {\tt musl} libc, unoptimized versions),
\item we do not reclaim -- and thus later re-allocate -- heap memory from terminated \fwp s, instead only zeroing it out, and using it to satisfy future {\tt eos\_sbrk} calls,
\item we reuse the threads active in each \fwp, instead only resetting their instruction pointer to the appropriate post-initialization execution point which has the side effect of avoiding thread allocation and scheduling overheads beyond suspending the thread.
\end{inparaenum}
These optimizations culminate in a system that can handle exceedingly high churn and scalability -- \fwp\ chain initialization converges on {\tt memcpy} overheads, and chain activation in response to a new client takes low 10s of microseconds.


\head{\fwp\ scheduling.}
We specialize the user-level scheduling policies within \eos\ to manage untrusted \fwp s that require low-latency computations.
The scheduling policy aims to prevent any \fwp\ from monopolizing the CPU, and from interfering with the progress of other \fwp s.
Additionally, as all scheduling operations represent overhead that can impact system throughput, they must be as rare and efficient as possible, while maintaining inter-\fwp\ isolation.
Given these goals, in the current work we focus on simplicity in the scheduling policy, and the careful usage of timer interrupts to balance each \fwp 's progress with scheduler overhead.

Each core separately schedules the \fwp s assigned to it using a fixed-priority, round-robin scheduling policy.
The quantum chosen to preempt an executing \fwp\ is specifically calibrated to enable the average \fwp\ to complete its execution cooperatively (thus avoiding timer overheads), and round-robin prevents starvation.
To implement this, user-level schedulers use the kernel's facilities to dispatch to a thread and pass the time that the next timer interrupt should fire.
We use modern x86 processor local-APIC support for specifying one-shot timer interrupts with cycle-accuracy (called ``TSC Deadline Timers'' in Intel documents).
Each scheduler receives messages from the \mma\ to activate its \fwp s.

The simplicity of the scheduling policy and our optimized use of timer interrupts, together enable the necessary efficiency for line-rate computations, while guaranteeing progress and performance predictabiliy in spite of the large-scale, multi-tenant environment.

\subsection{Message Pool Management}
\label{ss:shm}

To support multi-tenancy, \fwp s provide isolation for local memory, CPU processing, and access to system resources.
However, message pool management provides both inter-\fwp\ isolation and coordination.

\head{Ring-buffers for {\em both} coordination {\em and} memory management.}
Each \fwp 's message pool is associated with two ring buffers that track {\em both} how to transmit and receive messages, {\em and} the allocation and deallocation of messages.
These ring buffers are similar to NIC DMA ring buffers.
However, unlike traditional driver ring buffers, \eos\ makes the observations that (1) general purpose memory allocation facilities ({\tt malloc/free}) can have significant overhead for high message arrival rates, and complicate the coordinated memory management between the \mma\ and \fwp s; and (2) the ring buffers are organized to track not only incoming and outgoing messages, but also free memory.

A reception ring buffer contains a set of references to message slots into which incoming data can be copied, and the transmission ring buffer contains references to messages to move downstream in the \fwp\ chain.
The \mma\ dequeues messages from an \fwp 's transmit ring, copies the data, and enqueues a message in the recipient's ring.
In this way, the \mma\ acts directly as a software DMA accelerator between \fwp s.
Each ring buffer entry has a set of bits that tracks the state of the entry: {\sf transmit} -- ready to send the message, {\sf receive} -- empty message to transmit into, {\sf ready} -- populated message ready for processing, {\sf free} -- ready to be reallocated by the \fwp, or {\sf unused} -- an unused ring buffer entry (with an ignored pointer).
Thus, the \mma\ transitions ring buffer entries in transmit rings from the {\sf transmit} to the {\sf free} state after copying the message, thus signaling the message's reused; and it transitions receive ring entries from {\sf receive} to {\sf ready} after copying data into the message.

Message pools are managed by \fwp s as a span of MTU-sized message slots, and unlike traditional NIC DMA ring buffers, the ring buffers include an entry for {\em each} message slot.
When a message arrives in a message pool, the \fwp\ dequeues it from its receive ring -- transitioning the ring entry from {\sf ready} to {\sf unused}, processes it, and later adds it to the transmission ring buffer -- transitioning the entry from {\sf unused} to {\sf transmit}.
\fwp s must maintain a sufficient number of messages in reception rings in the {\sf receive} state to compensate for the scheduling latencies due to multiplexing the CPU among many \fwp s.
Thus, after it finishes processing pending messages, it will move freed messages from the transmit ring ({\sf free} $\to$ {\sf unused}), into the reception ring ({\sf unused} $\to$ {\sf receive}).
In this way, message {\em liveness} is managed indirectly through the ring buffers.

\head{Message pools and isolation.}
The ring buffer design decouples the {\em message memory} from the {\em meta-data} to coordinate the data movement and liveness between \fwp s and the \mma.
In doing so, \eos\ avoids lock-based protection of the rings, instead relying on wait-free mechanisms that guarantee execution progress of both \fwp s and the \mma.
This has the benefit of minimizing coherency overheads in ring coordination, and avoiding critical sections which threaten the \mma 's starvation.
Additionally, it enables \fwp s to have more restrictive access rights to the pool than the ring buffer, for example, providing integrity by mapping the pool read-only.


\subsection{Memory Movement Accelerator}
\label{ss:mma}

Our initial experiments showed that naively copying packets between stages in a DPDK-based NFV pipeline decreased throughput by more than 50\%.
However, we also found that a core devoted to data movement has a throughput of around 30 Gb/s, which is sufficient for line-rate.
By using the parallelism of the underlying processor and specializing cores to run the \mma, we achieve both isolation and high throughput by taking message movement out of the critical path.

The \mma\ has read-write access to all message pools.
It maintains a mapping between both pairs of transmit and receive ring buffers, and their associated pools, and continuously iterates through all such pairs, transferring messages when it finds a transmission.
The \mma\ provides two essential services: data-movement by copying transmitted messages, and event notification of the receiving \fwp s.
The \mma 's \fwp\ event notification is efficient as it simply sends a message to the scheduler controlling the \fwp 's core.
Though the current system uses only a single \mma, more cores can be devoted to this, should it require more memory movement throughput in the future.

\head{\mma\ optimizations.}
The \mma\ is on the data-path of all \fwp\ interactions, including message reception, thus it must be able to move messages at faster than line rate.
The \mma\ iterates through all \fwp\ transmit rings, and (1) copies data between message pools while updating rings, and (2) activates the downstream \fwp\ by sending an event (through a ring buffer) to the scheduler on that \fwp 's core.
The data-structures linking transmit and reception rings are laid out in an array to leverage the processor's prefetcher as the \mma\ iterates over them.
The initial implementation of the operations on the ring buffers were straight-forward, but cache-coherency overheads, possibly for each ring entry, hurt throughput.
To address this, we added two optimization:
\begin{itemize}
\item Double-cache-line (128B) {\em caches} are added to both the enqueue and dequeue operations.
      These caches are in local memory outside of the ring, thus their modifications are free of coherency traffic.
      Transmitting a message adds it to the transmit queue cache, and only when it is full is it flushed to the ring buffer.
      This batches what would be eight separate ring updates into essentially a single {\tt memcpy} of 128 bytes.
      To avoid cached entries that are not yet transferred into the ring from having delayed (or starved) processing, when an \fwp\ has completed processing, and is going to block, it flushes its cache to its transmit ring buffer.
      Similarly, when the ring buffers are dequeued, entries are copied out into a double-cache-line cache, and subsequent accesses first check the cache.
      The caches are 128B to match the Intel policy of fetching double-cache-lines at a time.
\item These caches enable messages to be viewed in batches.
      This enables a second optimization to use explicit software prefetch instructions to load all referenced messages into the core's cache.
      This optimization is particularly effective as the processing of the messages is temporally proximate.
\item Naming of different messages uses direct virtual addresses.
      Though the \mma\ is isolated from \fwp s, they share a single virtual address space~\cite{chase92opal,druschel93fbufs}.
      To maintain protection, all local memory for both the \mma\ and each \fwp\ is isolated and uses overlapping address, and when the \mma\ and \fwp s pass a message, they validate that it lies within the message pool's boundaries.
\end{itemize}
These optimizations contribute to \eos ' high message throughput.
However, should they be insufficient due to 
too many \fwp s or long chains, the \mma\ can trivially partition the ring buffers, thus scale to multiple cores.

\vspace{-2mm}
\subsection{Network Interface Integration}
\label{ss:dpdk}

\eos\ uses DPDK for direct access to the NIC via kernel-bypass.
Our port of DPDK to \eos\ is conducted mainly as a new Environment Abstraction Layer (EAL), thus minimizing the impact on the DPDK code-base.
DPDK transmits and receives packets via the \mma, but, unlike other \fwp s, it has a number of heightened privileges.
First, DPDK is used in poll-mode, and we devote a core to polling for and receiving packets, and another to transmitting them.

\head{Packet reception.}
Incoming packets are demultiplexed to their corresponding \fwp s via the flow mapping facilities in DPDK.
In this way, \eos\ has mechanisms to maintain the mappings of IPs and ports to specific \fwp-chains, but we leave the policy of creating those mappings to a cluster manager such as Kubernetes or a Software-Defined Networking (SDN) controller.
If a flow maps to an \fwp-chain that is not yet active, a chain is retrieved from the \fwp-chain cache, and activated.
The \fwp-chain cache is populated with \fwp\ chains by the \fwp\ manager.

DPDK packet pools are treated as \eos\ message pools, and the \mma\ copies packets into downstream \fwp s.
Intelligent hardware with flow direction built in might enable zero-copy here~\cite{sharma17flexnic}, and \eos\ could be modified to use this support in the future.
Flows that map to an active \fwp-chain are placed in a message pool transmit ring buffer, and the \mma\ copies the data accordingly.

\head{Packet transmission.}
A final optimization avoids a packet copy on the transmit path.
When the last \fwp\ in the chain transmits to DPDK, the \mma\ omits the copy, and instead enables DPDK to add a direct reference to the packet to its own DMA ring buffers.
Later, when the NIC signals the successful transmission of the packet, DPDK signals the transmission to the message pool so that the packet can be reclaimed.

\begin{figure*}[!ht]
\begin{center}
\includegraphics[trim={0cm 0cm 0cm 0cm}]{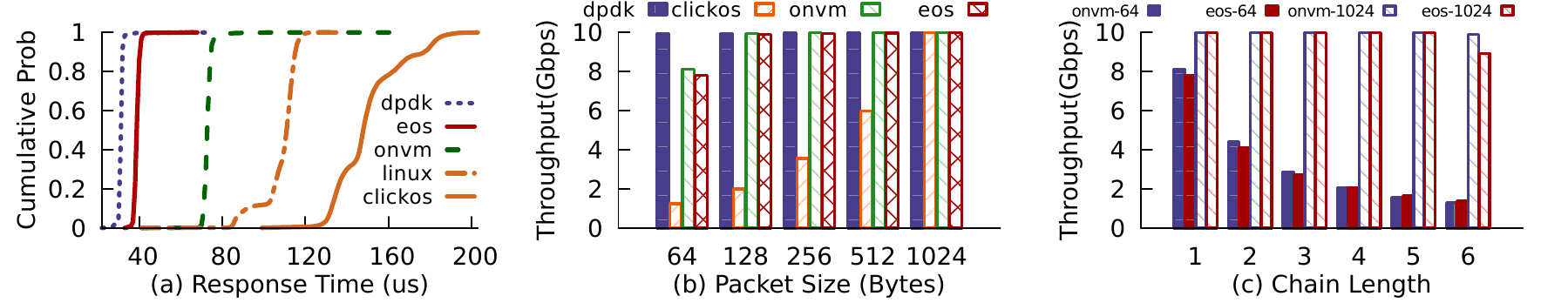}
\end{center}
\vspace{-1.2em}
\caption{{\small
(a) \eos provides substantially better latency, and reduced jitter compared to Linux processes and NFV platforms like OpenNetVM and ClickOS.
(b) Throughput of each system with different packets sizes.
(c) \eos provides isolation and adds negligible overheads compared to OpenNetVM (no isolation) for different chain length for messages of size 64 and 1024 bytes.
}}
\label{fig:micro-bench}
\vspace{-1em}
\end{figure*}

%% file: sections/eval.tex







All experiments are run on CloudLab Wisconsin c220g1 series nodes.
These are two socket, 8 core, Intel(R) Xeon(R) CPU E5-2630 v3 @ 2.40GHz Intel processors with 128GB ECC Memory (8x 16 GB DDR4 1866 MHz dual rank RDIMMs).
Systems are connected via Dual-port Intel X520-DA2 10Gb NIC (PCIe v3.0, 8 lanes) networking cards.
For \eos\ we use less than 1GB of the system's memory due to the underlying $\mu$-kernel's 32-bit address space limitations.

\subsection{Latency and Throughput}
\label{ss:micro}

We first evaluate the latency and performance predictability of \eos\ compared to other high performance networking platforms.
Figure~\ref{fig:micro-bench}(a) shows the response time distribution (in microseconds) for an ICMP ping response Click~\cite{kohler00click} element implemented as either: a DPDK process, an OpenNetVM NF (ONVM), a standard linux process with kernel-based IO, a ClickOS NF in a Xen VM, or an \fwp\ in \eos.
The results show that \eos\ significantly outperforms all of these techniques (by up to 3.8X in average latency), except for DPDK.
DPDK is slightly better because it can run only a single service at a time and thus does not need to copy packets from the initial receive DMA ring to a separate pool.
In contrast, \eos\ provides a platform to potentially run thousands of distinct services, and thus needs to offer stronger isolation via copying.

Figure~\ref{fig:micro-bench}(b) shows the maximum throughput of different approaches when forwarding traffic from pktgen, a high speed packet generator.
\eos\ again provides better performance than ClickOS, while offering stronger isolation than DPDK and ONVM, which rely on globally shared memory pools for zero-copy IO.

Next we evaluate the performance of \eos communication by comparing with ONVM. We run a chain of NFs on the same core that each forward small (64B) or big (1024B) packets, thus both systems have context switch overhead by passing a packet to the next NF.
In addition, \eos has copying overhead from the \mma\ to enforce isolation.
The results in Figure~\ref{fig:micro-bench}(c), show that as the chain length increases, the throughput of 64B packet drops for both \eos and ONVM affected by different overheads. 
The main overhead of \eos is data copying, while the overhead of Linux context switches and scheduling dominates ONVM.
When the chain length is smaller than 3, the overhead of copying is less than 8\%, and \eos outperforms ONVM when the chain is longer as the Linux system overheads increase.
The throughput with 1024B packets maintains line rate for both systems when the chain length is smaller than 6, but \eos sees a throughput decrease when the chain is longer as one MMA is not able to handle copies for all FWPs.




\begin{figure*}[tb]
\begin{center}
\includegraphics[trim={0cm 0cm 0cm 0cm}]{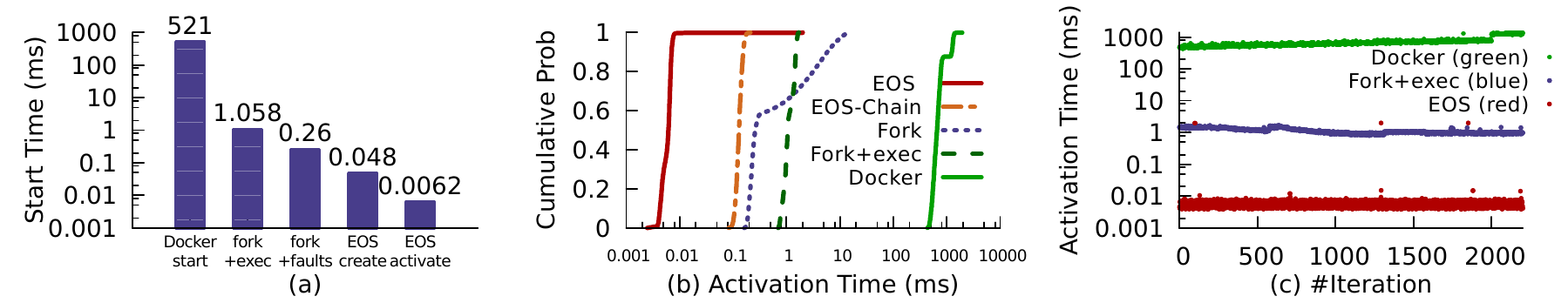}
\end{center}
\vspace{-1.2em}
\caption{
{\small \eos\ provides orders of magnitude better startup time than other approaches and does not suffer from scalability problems when starting larger numbers of \fwp s.}
}
\label{fig:fwp-init}
\vspace{-1em}
\end{figure*}

\subsection{Startup Time}
\label{ss:eval-boot}

\head{FWP Initialization and Activation.}
In Linux, initializing a process involves calling {\tt fork} (and possibly {\tt execve}).
For Docker containers, a {\tt docker run} command is similar, but includes additional system calls to configure namespaces and maintain container metadata.
In order to optimize the fast path of readying a cached \fwp, \eos\ separates out creation from activation.
For \eos, creation involves transitioning from the Object File to Cached state in Figure~\ref{fig:lifecycle}, including setting up page tables, capability tables, and thread creation.
We record the start time for 10,000 iterations of starting a container, process, or \fwp\ and report the median in Figure~\ref{fig:fwp-init} (a).
Note the log scale; we use median time values since as described below, Container creation becomes more slowly over time so the average is skewed by these outliers.
We compare against two variants of Linux processes: "fork + exec" loads a different binary whereas "fork + faults" mimics loading the service's working set by issuing writes to 8 different pages to trigger page faults.
These approaches are 5-20X slower than the comparable "EOS create" approach (dashed lines in Figure~\ref{fig:lifecycle}).

Once an \fwp\ has been created, \eos\ keeps copies of it in a cache which can be quickly activated on demand (solid lines in Figure~\ref{fig:lifecycle}).
Cached activation improves \eos performance by another order of magnitude, allowing new processing entities to be instantiated in 6.2 microseconds.
Figure~\ref{fig:fwp-init}(b) presents a CDF of these approaches, including the activation cost for starting a full chain of 10 \fwp s, which remains an order of magnitude faster than fork+exec.

\head{FWP Scalability}.
Further, we have found that containers suffer from poor scalability -- as the number of containers rise, the start time worsens.
Similar behavior has been shown previously for virtual machines~\cite{manco17lightvm}.
In Figure~\ref{fig:fwp-init}(c) we show the time to start a new container, exec a process, or activate an \fwp , when up to 2200 are started incrementally.
The Container case gradually drifts upward before hitting a step after 2000 containers.
The cost of starting the last container is 1.368 seconds versus 0.467 seconds for the first.
The standard deviation for containers is 236 ms versus only 0.08 ms for \fwp\ activation.
As long as sufficient \fwp s are available in the template cache, \eos\ provides nearly constant start time regardless of scale; if additional templates are needed, the \fwp\ manager can created them in parallel to the data path.
The \eos\ timeline has a few outlier points (11 out of 15K measurements are at 2ms), which we believe to be Non-Maskable Interrupts, or a bug in our scheduling logic.

\vspace{-1mm}
\subsection{Isolation}



\begin{figure}[!ht]
\vspace{-1.5em}
\begin{center}
\includegraphics[height=1.4in, trim={0cm 0cm 0cm 0cm}]{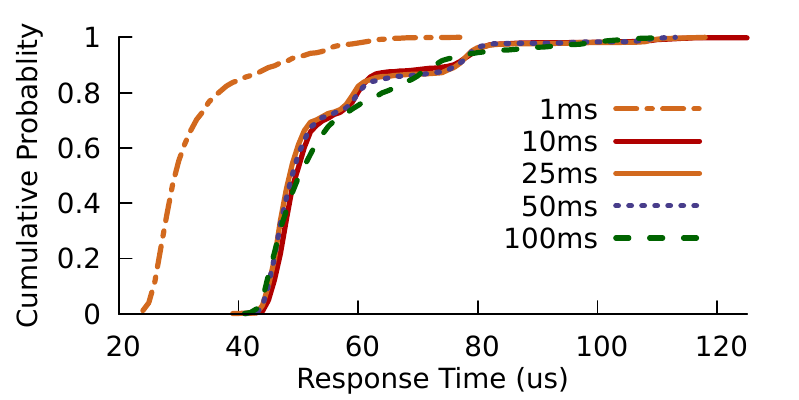}
\end{center}
\vspace{-1em}
\caption{
\eos\ just in time service instantiation for mobile clients with varying client inter-arrival rates.
}
\label{fig:churn}
\vspace{-1em}
\end{figure}

\head{Just in Time Service Instantiation.} To evaluate the impact of client churn in edge environments, we mimic an experiment from the LightVM paper~\cite{manco17lightvm}.
Clients send requests to an \eos\ based service at a configurable interval, and we assume that each new client request requires its own \fwp\ to be instantiated.
The new \fwp\ receives the incoming packet, produces a reply, and then terminates, representing a worst case churn scenario.
Figure~\ref{fig:churn} shows a response time CDF for \eos\ under different client arrival patterns.
The results show that even when a new client arrives every millisecond, 90\% of requests are serviced within 50 microseconds
Although we have not been able to successfully run the LightVM software on our testbed, we note that their paper produced a 90th percentile response time of 20 milliseconds (more than 400X worse) with clients arriving 10 times less frequently (10ms interval).
The \eos\ performance advantage comes from our extremely lightweight \fwp\ abstraction and our template cache that allows nearly instant instantiation.

\begin{figure}[!ht]
\vspace{-1.5em}
\begin{center}
\includegraphics[width=3in, trim={0cm 0cm 0cm 0cm}]{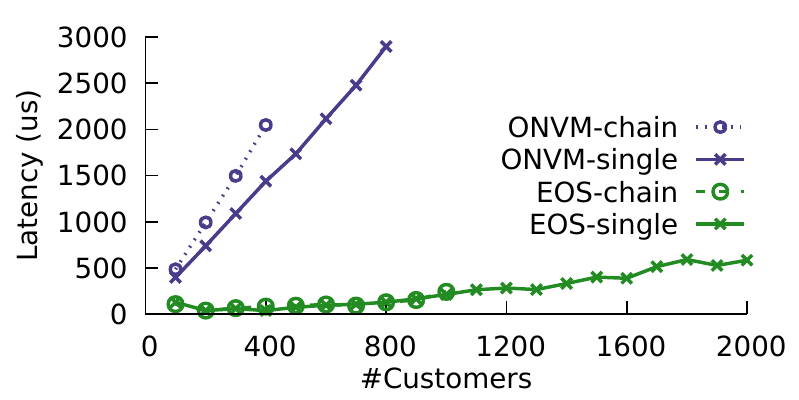}
\end{center}
\vspace{-1.2em}
\caption{\small Routing and processing latency for routing {\tt netperf} traffic for an increasing number of clients.
}
\label{fig:customer}
\vspace{-1.2em}
\end{figure}

\begin{figure*}[!ht]
        \centering
        \includegraphics[width=6.5in]{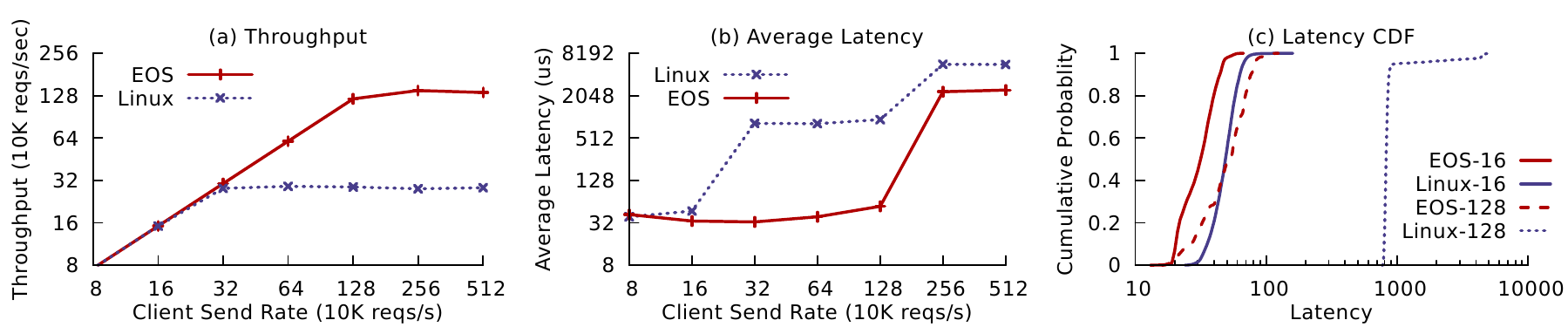}
        \vspace{-1em}
        \caption{\small Single memcached instance on one core.
        }
        \label{fig:single_mc}
        \vspace{-1em}
\end{figure*}

\begin{figure*}[!t]
        \vspace{-0.5em}
        \centering
        \includegraphics[width=6.5in]{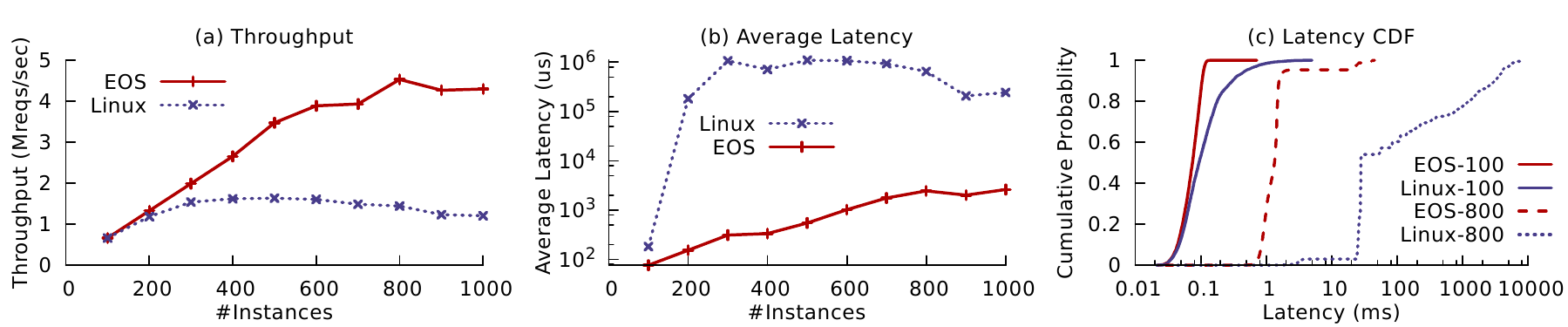}
        \vspace{-1em}
        \caption{\small Multiple memcached instances (1 per client) on 16 cores.
        }
        \label{fig:multi_mc}
        \vspace{-1em}
\end{figure*}

\head{Multi-Tenancy and Customer Isolation.}
An important job of edge-cloud systems, is acting as a middlebox to route a subset of requests to the cloud.
Figure~\ref{fig:customer} depicts the processing latency of processing and routing requests between {\tt netperf} client and server machines for an increasing number of concurrent clients.
We use three nodes, two running {\tt netperf} clients and servers, and the third running \eos\ or ONVM in the middle to act as the middlebox.
The systems run either a single firewall to filter flows or a 2 \fwp\ chain of firewall plus monitor, all implemented in Click, to further maintain statistics about flows.
Each customer is serviced by its own separate firewall or chain, thus is isolated from each other.
We measure the middlebox latency overhead (i.e., the added cost versus direct client/server connections from Figure~\ref{fig:scale_motiv}) as we increase the number of clients, and thus number of \fwp s (\eos) and Network Functions (NFs in ONVM).

Though ONVM is a highly optimized middlebox infrastructure, it relies on containers and expensive coordination mechanisms between NFs and the management layer.
Because of this, ONVM cannot scale past around 820 containers or 410 chains, and the added latency rises quickly with each new client.
\fwp s enable the system to scale past 2000 customers with an average increase in the latency of only around 0.3$\mu s$ per additional client.
Chaining in \eos\ adds negligible latency overhead thanks to our efficient scheduler notification and context switch, while ONVM sees an increasing gap since it relies on Linux's more heavyweight futexes and its underlying scheduling.

\subsection{Memcached}
\label{ss:eval-memcached}
Finally, we evaluate how \eos\ can provide a platform for low latency endpoint applications.
We implement an \fwp\ capable of parsing memcached UDP requests and use it to replace the standard socket interface in the memcached server.
The EOS controller can then be used to map incoming requests either to a single memcached \fwp\ (e.g., representing an edge cloud data cache) or one \fwp\ per client (e.g., representing private data stores for edge-connected IoT devices).
We compare \eos\ against Linux, either using a single memcached server or multiple.  Our workload, inspired by~\cite{nishtala_scaling_2013}, uses 135 byte value sizes and a 95\% get, 5\% set request mix generated by the mcblaster client.


\head{Single instance.} Figure~\ref{fig:single_mc} shows the throughput and latency when all clients connect to a single memcached server instance pinned to one core.
We use a 16-core server as the client, running one mcblaster process per core to ensure the client will not be the bottleneck.
Each client process sends requests at a configurable rate and we report the aggregate throughput and average latency of successful requests (i.e., dropped requests do not impact latency).
From Figure~\ref{fig:single_mc} we see that \eos\ can support a throughput of up to 1.4 million requests per second, a nearly 5X increase compared to Linux.
The response time of \eos\ is also substantially lower than Linux, and it can handle 8X the client request rate before seeing an increase in latency.
With very low client request rates, both systems perform similarly because \eos cannot take advantage of batching.
Since Linux is not able to keep up, it drops a large number of requests, e.g., 5.2\% at a 320K req/sec client rate.
In contrast, \eos\ does not see any requests drops at a 1.2M req/sec client rate.


\head{Multiple instance.} We next run a scalability test where each client is paired with its own memcached instance, either running as a Linux process or an \fwp.
The Linux processes are started in advance, whereas the \fwp s must be activated from the cache for the first request from a client.
Each client sends at a fixed rate of 10 Mbit/s and they are distributed across four hosts to prevent them being the bottleneck.
We distribute the memcached server instances evenly across the available cores of the host server -- for Linux all 16 cores are available, whereas for \eos\ only 12 cores are used for running \fwp s and 4 are used for the system services.
As we increase the number of clients, the aggregate request rate rises, with Linux hitting its peak throughput with 300 memcached clients and servers.
\eos\ is able to scale substantially further, hitting a maximum throughput over 4M req/sec with 800 clients.
The Linux server, overwhelmed with the number of memcached processes, has a response rate of nearly 1 second, whereas \eos\ maintains an average latency below 1 millisecond for up to 600 memcached instances.
From the latency CDF, we observe that even with only 100 memcached instances, Linux has much higher tail latency than \eos, and that with 800 instances Linux has more than three orders of magnitude worse tail latency.
Keep in mind that these latency metrics ignore dropped, requests -- with 800 instances, \eos\ drops 13\% of requests, whereas Linux drops 66\%.

\head{TCP vs UDP.} Our memcached implementation is based on UDP since \eos\ does not provide a TCP stack.
Adding a high performance DPDK-based TCP stack~\cite{jeong_mtcp:_2014} would be straightforward, and we expect the performance difference between Linux and \eos\ to grow even larger in this case.
In the current UDP implementation, the high drop rate seen in Linux has no impact on its throughput or latency, but with TCP this would trigger congestion control and retransmissions, leading to even worse performance.

%% file: sections/related.tex
\head{Scalable multi-tenant isolation.}
Significant research addresses the increasing {\em churn} seen in serverless computing~\cite{lagar_cavilla09snowflock,madhavapeddy15jitsu,nitu_swift_2017,manco17lightvm} by decreasing the startup and teardown costs of virtual machines.
Light-weight systems such as unikernels~\cite{manco17lightvm,madhavapeddy13unikernels} only further increase the agility of these systems.
In contrast, \eos\ is motivated by the potentially enormous churn and large-scale isolation requirements of the edge cloud, providing service to transient mobile and IoT devices.
The \fwp\ abstraction, and activation based on the \fwp-chain cache provide low-overhead isolation and message pools for effective communication that can handle the unprecedented churn.
Denali~\cite{whitaker02denali} separates the protection provided by a VMM from the abstractions within the VM, and enables lightweight VM contexts that scale from tens to low hundred's of VMs.
\eos\ focuses on extremely fast \fwp\ activation times for on-the-fly instantiation, and \mma -coordinated communication through chains of \fwp s to enable service composition from multiple tenants.
Multiple projects have increased the efficiency of containers by specializing the environment for more efficient boot-up.
{\sf Cntr}~\cite{thalheim18cntr} includes only the application-specific context in a container, while {\sf SOCK}~\cite{oakes18sock} specializes the container to use efficient kernel operations, and uses a Zygote mechanism paired with a cache to accelerate container creation for stateless computations.
For isolated edge computation instantiation, \eos\ compares favorably to forking of minimal Linux processes (two orders of magnitude faster start-time) which is the lower-bound for many such techniques.
These projects have startup latencies in the milliseconds versus \fwp s in the 10s of microseconds.
Additionally due to \fwp\ optimizations, \eos\ also maintains significantly lower edge application latencies at scale than Linux (100s of $\mu$-seconds vs. a second for {\tt memcached}).

\head{Lightweight isolation.}
Wedges~\cite{bittau08wedge}, Light-weight Contexts~\cite{litton16lwcontexts}, and SpaceJMP~\cite{elhajj16spacejmp} expand the UNIX interface to include lightweight facilities for controlling and changing protection domains.
Similarly, Dune~\cite{belay12dune} uses hardware virtualization support to provide user-level control over page-tables, and dIPC~\cite{vilanova17dipc} uses hardware support to bypass the kernel during inter-protection domain communication.
\eos\ instead relies on a highly-optimized $\mu$-kernel's core support for secure and efficient control flow management, protection domains, and capability-based access control.
We target abstractions to support immense churn rates, and efficient communication with complete isolation via the \mma.
To efficiently use the limited resources in the edge cloud, \eos\ leverages this support to scale to more than two thousand \fwp s in less than 1GB of RAM while maintaining line-rate communication.

\head{User-level, high-performance networking.}
User-level network processing has long been proposed~\cite{eicken95unet} to better utilize HW and reach line-rate throughput.
Isolation is provided when paired with the early demultiplexing of networking packets~\cite{tennenhouse89layers_harmful,engler95exokernel}.
Shared memory for zero-copy communication, and batched processing have pushed these techniques into Gb-level networking~\cite{belay14ix,rizzo12netmap,han2012megapipe}.
DPDK and other kernel by-pass techniques have also pushed middlebox network function processing effectively into VMs~\cite{martins14clickos}, and containers~\cite{zhang_opennetvm:_2016}.
\eos\ expands on these techniques by integrating them with large-scale multi-tenancy via the \mma, and the strong isolation of \fwp s.
NetBricks~\cite{panda16netbricks} implement network processing functions in a memory-safe language (Rust), thus relying on the software isolation in a single thread.
\eos\ effectively uses the parallelism of the underlying hardware, and the \mma\ to maintain memory safety, but also provides {\em temporal isolation} by executing all \fwp s in separate threads that are explicitly scheduled by the run-time.

\head{Hardware NIC demultiplexing.}
Hardware-based early demultiplexing of networking packets has enabled isolated, high-performance library-based system services~\cite{peter15arrakis}.
While this avoids the use of shared memory pools, it relies on network hardware support for multiple queues to isolate principals.
Such support is limited, e.g., the common Intel 82599 chipset for 10Gbps NICs only supports up to 128 queues~\cite{noauthor_82599-10-gbe-controller-brief.pdf_nodate}.
Intelligent NICs take this idea further by supporting demultiplexing with higher fidelity~\cite{sharma17flexnic}.
\eos\ supports a high level of scalability required for the multi-tenant edge cloud, thus uses software techniques to safely demultiplex packets by devoting cores to act as \mma s without relying on specialized hardware.
Results show that the system can maintain line-rate despite using these software accelerators.

%% file: sections/conc.tex
The increasing prevalence of mobile computations and the Internet of Things requires both scalable isolation facilities for multi-tenancy in the edge, and the agility to handle high churn.
This paper has described \eos, an OS for edge cloud computation that introduces a Feather-Weight Process abstraction for low-overhead isolation that is paired with a cache of post-initialization checkpointed \fwp -chains to provide the microsecond scale activation times necessary to handle high churn.
Isolation is facilitated with a specialized core devoted to accelerating moving messages between \fwp s, thus maintaining isolation.

We show that \eos\ provides more than a 3.8X reduction in ping latency and more than 2X throughput increase compared to ClickOS -- a system that also provides isolated computation -- for middlebox computations.
More importantly, \eos\ can create \fwp s for client computation in 25-50 microseconds, even when they are created every millisecond, and can scale to over 2000 \fwp s while maintaining low latency, even with a very limited amount of memory.
For edge applications like memcached, \eos\ has more than three orders of magnitude decreases in latency when running over 300 server instances simultaneously. 
We believe that \eos\ paves the way for closely integrating the edge cloud into -- and augmenting the capabilities of -- the increasing prevalence of mobile and embedded devices.

%% file: ms.bbl

\begin{thebibliography}{50}


\ifx \showCODEN    \undefined \def \showCODEN     #1{\unskip}     \fi
\ifx \showDOI      \undefined \def \showDOI       #1{#1}\fi
\ifx \showISBNx    \undefined \def \showISBNx     #1{\unskip}     \fi
\ifx \showISBNxiii \undefined \def \showISBNxiii  #1{\unskip}     \fi
\ifx \showISSN     \undefined \def \showISSN      #1{\unskip}     \fi
\ifx \showLCCN     \undefined \def \showLCCN      #1{\unskip}     \fi
\ifx \shownote     \undefined \def \shownote      #1{#1}          \fi
\ifx \showarticletitle \undefined \def \showarticletitle #1{#1}   \fi
\ifx \showURL      \undefined \def \showURL       {\relax}        \fi
\providecommand\bibfield[2]{#2}
\providecommand\bibinfo[2]{#2}
\providecommand\natexlab[1]{#1}
\providecommand\showeprint[2][]{arXiv:#2}

\bibitem[\protect\citeauthoryear{Banga, Druschel, and Mogul}{Banga
  et~al\mbox{.}}{1999}]%
        {banga99resourceCont}
\bibfield{author}{\bibinfo{person}{Gaurav Banga}, \bibinfo{person}{Peter
  Druschel}, {and} \bibinfo{person}{Jeffrey~C. Mogul}.}
  \bibinfo{year}{1999}\natexlab{}.
\newblock \showarticletitle{Resource containers: a new facility for resource
  management in server systems}. In \bibinfo{booktitle}{\emph{OSDI '99:
  Proceedings of the third symposium on Operating systems design and
  implementation}}. \bibinfo{publisher}{USENIX Association},
  \bibinfo{address}{Berkeley, CA, USA}, \bibinfo{pages}{45--58}.
\newblock
\showISBNx{1-880446-39-1}


\bibitem[\protect\citeauthoryear{Belay, Bittau, Mashtizadeh, Terei,
  Mazi\`{e}res, and Kozyrakis}{Belay et~al\mbox{.}}{2012}]%
        {belay12dune}
\bibfield{author}{\bibinfo{person}{Adam Belay}, \bibinfo{person}{Andrea
  Bittau}, \bibinfo{person}{Ali Mashtizadeh}, \bibinfo{person}{David Terei},
  \bibinfo{person}{David Mazi\`{e}res}, {and} \bibinfo{person}{Christos
  Kozyrakis}.} \bibinfo{year}{2012}\natexlab{}.
\newblock \showarticletitle{Dune: Safe User-level Access to Privileged CPU
  Features}. In \bibinfo{booktitle}{\emph{Proceedings of the 10th {USENIX}
  Symposium on Operating Systems Design and Implementation (OSDI'12),
  Hollywood, CA, USA, October 8-10}}.
\newblock


\bibitem[\protect\citeauthoryear{Belay, Prekas, Klimovic, Grossman, Kozyrakis,
  and Bugnion}{Belay et~al\mbox{.}}{2014}]%
        {belay14ix}
\bibfield{author}{\bibinfo{person}{Adam Belay}, \bibinfo{person}{George
  Prekas}, \bibinfo{person}{Ana Klimovic}, \bibinfo{person}{Samuel Grossman},
  \bibinfo{person}{Christos Kozyrakis}, {and} \bibinfo{person}{Edouard
  Bugnion}.} \bibinfo{year}{2014}\natexlab{}.
\newblock \showarticletitle{IX: A Protected Dataplane Operating System for High
  Throughput and Low Latency}. In \bibinfo{booktitle}{\emph{Proceedings of the
  11th USENIX Conference on Operating Systems Design and Implementation
  (OSDI)}}.
\newblock


\bibitem[\protect\citeauthoryear{Bittau, Marchenko, Handley, and Karp}{Bittau
  et~al\mbox{.}}{2008}]%
        {bittau08wedge}
\bibfield{author}{\bibinfo{person}{Andrea Bittau}, \bibinfo{person}{Petr
  Marchenko}, \bibinfo{person}{Mark Handley}, {and} \bibinfo{person}{Brad
  Karp}.} \bibinfo{year}{2008}\natexlab{}.
\newblock \showarticletitle{Wedge: Splitting Applications into
  Reduced-privilege Compartments}. In \bibinfo{booktitle}{\emph{Proceedings of
  the 5th USENIX Symposium on Networked Systems Design and Implementation
  (NSDI)}}.
\newblock


\bibitem[\protect\citeauthoryear{Chase, Baker-Harvey, Levy, and Lazowska}{Chase
  et~al\mbox{.}}{1992}]%
        {chase92opal}
\bibfield{author}{\bibinfo{person}{Jeffrey~S. Chase}, \bibinfo{person}{Miche
  Baker-Harvey}, \bibinfo{person}{Henry~M. Levy}, {and}
  \bibinfo{person}{Edward~D. Lazowska}.} \bibinfo{year}{1992}\natexlab{}.
\newblock \showarticletitle{Opal: A Single Address Space System for 64-Bit
  Architectures}.
\newblock \bibinfo{journal}{\emph{Operating Systems Review}}
  \bibinfo{volume}{26}, \bibinfo{number}{2} (\bibinfo{year}{1992}),
  \bibinfo{pages}{9}.
\newblock
\urldef\tempurl%
\url{citeseer.ist.psu.edu/58003.html}
\showURL{%
\tempurl}


\bibitem[\protect\citeauthoryear{Dennis and Horn}{Dennis and Horn}{1983}]%
        {dennis66capabilities}
\bibfield{author}{\bibinfo{person}{Jack~B. Dennis} {and} \bibinfo{person}{Earl
  C.~Van Horn}.} \bibinfo{year}{1983}\natexlab{}.
\newblock \showarticletitle{Programming semantics for multiprogrammed
  computations}.
\newblock \bibinfo{journal}{\emph{Commun. ACM}} \bibinfo{volume}{26},
  \bibinfo{number}{1} (\bibinfo{year}{1983}), \bibinfo{pages}{29--35}.
\newblock
\showISSN{0001-0782}
\urldef\tempurl%
\url{https://doi.org/10.1145/357980.357993}
\showDOI{\tempurl}


\bibitem[\protect\citeauthoryear{Docker}{Docker}{2018}]%
        {docker}
Docker \bibinfo{year}{2018}\natexlab{}.
\newblock \bibinfo{title}{Docker: https://www.docker.com/}.
\newblock
\newblock


\bibitem[\protect\citeauthoryear{Dragovic, Fraser, Hand, Harris, Ho, Pratt,
  Warfield, Barham, and Neugebauer}{Dragovic et~al\mbox{.}}{2003}]%
        {dragovic03xen}
\bibfield{author}{\bibinfo{person}{B. Dragovic}, \bibinfo{person}{K. Fraser},
  \bibinfo{person}{S. Hand}, \bibinfo{person}{T. Harris}, \bibinfo{person}{A.
  Ho}, \bibinfo{person}{I. Pratt}, \bibinfo{person}{A. Warfield},
  \bibinfo{person}{P. Barham}, {and} \bibinfo{person}{R. Neugebauer}.}
  \bibinfo{year}{2003}\natexlab{}.
\newblock \showarticletitle{Xen and the Art of Virtualization}. In
  \bibinfo{booktitle}{\emph{Proceedings of the ACM Symposium on Operating
  Systems Principles (SOSP)}}.
\newblock


\bibitem[\protect\citeauthoryear{Druschel and Peterson}{Druschel and
  Peterson}{1993}]%
        {druschel93fbufs}
\bibfield{author}{\bibinfo{person}{Peter Druschel} {and}
  \bibinfo{person}{Larry~L. Peterson}.} \bibinfo{year}{1993}\natexlab{}.
\newblock \showarticletitle{Fbufs: A High-Bandwidth Cross-Domain Transfer
  Facility}. In \bibinfo{booktitle}{\emph{Symposium on Operating Systems
  Principles}}. \bibinfo{pages}{189--202}.
\newblock


\bibitem[\protect\citeauthoryear{El~Hajj, Merritt, Zellweger, Milojicic,
  Achermann, Faraboschi, Hwu, Roscoe, and Schwan}{El~Hajj
  et~al\mbox{.}}{2016}]%
        {elhajj16spacejmp}
\bibfield{author}{\bibinfo{person}{Izzat El~Hajj}, \bibinfo{person}{Alexander
  Merritt}, \bibinfo{person}{Gerd Zellweger}, \bibinfo{person}{Dejan
  Milojicic}, \bibinfo{person}{Reto Achermann}, \bibinfo{person}{Paolo
  Faraboschi}, \bibinfo{person}{Wen-mei Hwu}, \bibinfo{person}{Timothy Roscoe},
  {and} \bibinfo{person}{Karsten Schwan}.} \bibinfo{year}{2016}\natexlab{}.
\newblock \showarticletitle{SpaceJMP: Programming with Multiple Virtual Address
  Spaces}. In \bibinfo{booktitle}{\emph{Proceedings of the Twenty-First
  International Conference on Architectural Support for Programming Languages
  and Operating Systems (ASPLOS)}}.
\newblock


\bibitem[\protect\citeauthoryear{Elphinstone and Heiser}{Elphinstone and
  Heiser}{2013}]%
        {elphinstone13l4}
\bibfield{author}{\bibinfo{person}{Kevin Elphinstone} {and}
  \bibinfo{person}{Gernot Heiser}.} \bibinfo{year}{2013}\natexlab{}.
\newblock \showarticletitle{From {L3 to seL4} what have we learnt in 20 years
  of {L4} microkernels?}. In \bibinfo{booktitle}{\emph{Proceedings of the 24th
  ACM Symposium on Operating Systems Principles (SOSP)}}.
\newblock


\bibitem[\protect\citeauthoryear{Engler, Kaashoek, and O'Toole}{Engler
  et~al\mbox{.}}{1995}]%
        {engler95exokernel}
\bibfield{author}{\bibinfo{person}{Dawson~R. Engler}, \bibinfo{person}{Frans
  Kaashoek}, {and} \bibinfo{person}{James O'Toole}.}
  \bibinfo{year}{1995}\natexlab{}.
\newblock \showarticletitle{Exokernel: An Operating System Architecture for
  Application-Level Resource Management}. In
  \bibinfo{booktitle}{\emph{Proceedings of the 15th ACM Symposium on Operating
  System Principles}}. \bibinfo{publisher}{ACM}, \bibinfo{address}{Copper
  Mountain Resort, Colorado, USA}, \bibinfo{pages}{251--266}.
\newblock


\bibitem[\protect\citeauthoryear{Gadepalli, Gifford, Baier, Kelly, and
  Parmer}{Gadepalli et~al\mbox{.}}{2017}]%
        {gadepalli17tcaps}
\bibfield{author}{\bibinfo{person}{Phani~Kishore Gadepalli},
  \bibinfo{person}{Robert Gifford}, \bibinfo{person}{Lucas Baier},
  \bibinfo{person}{Michael Kelly}, {and} \bibinfo{person}{Gabriel Parmer}.}
  \bibinfo{year}{2017}\natexlab{}.
\newblock \showarticletitle{Temporal Capabilities: Access Control for Time}. In
  \bibinfo{booktitle}{\emph{Proceedings of the 38th IEEE Real-Time Systems
  Symposium}}.
\newblock


\bibitem[\protect\citeauthoryear{Gupta, Cherkasova, Gardner, and Vahdat}{Gupta
  et~al\mbox{.}}{2006}]%
        {gupta_enforcing_2006}
\bibfield{author}{\bibinfo{person}{Diwaker Gupta}, \bibinfo{person}{Ludmila
  Cherkasova}, \bibinfo{person}{Rob Gardner}, {and} \bibinfo{person}{Amin
  Vahdat}.} \bibinfo{year}{2006}\natexlab{}.
\newblock \showarticletitle{Enforcing {Performance} {Isolation} {Across}
  {Virtual} {Machines} in {Xen}}. In \bibinfo{booktitle}{\emph{Proceedings of
  the {ACM}/{IFIP}/{USENIX} 2006 {International} {Conference} on {Middleware}}}
  \emph{(\bibinfo{series}{Middleware '06})}.
  \bibinfo{publisher}{Springer-Verlag New York, Inc.}, \bibinfo{address}{New
  York, NY, USA}, \bibinfo{pages}{342--362}.
\newblock
\urldef\tempurl%
\url{http://dl.acm.org/citation.cfm?id=1515984.1516011}
\showURL{%
\tempurl}


\bibitem[\protect\citeauthoryear{Han, Jang, Panda, Palkar, Han, and
  Ratnasamy}{Han et~al\mbox{.}}{2015}]%
        {han_softnic:_2015}
\bibfield{author}{\bibinfo{person}{Sangjin Han}, \bibinfo{person}{Keon Jang},
  \bibinfo{person}{Aurojit Panda}, \bibinfo{person}{Shoumik Palkar},
  \bibinfo{person}{Dongsu Han}, {and} \bibinfo{person}{Sylvia Ratnasamy}.}
  \bibinfo{year}{2015}\natexlab{}.
\newblock \bibinfo{booktitle}{\emph{{SoftNIC}: {A} {Software} {NIC} to
  {Augment} {Hardware}}}.
\newblock \bibinfo{type}{{T}echnical {R}eport} UCB/EECS-2015-155.
  \bibinfo{institution}{EECS Department, University of California, Berkeley}.
\newblock
\urldef\tempurl%
\url{http://www.eecs.berkeley.edu/Pubs/TechRpts/2015/EECS-2015-155.html}
\showURL{%
\tempurl}


\bibitem[\protect\citeauthoryear{Han, Marshall, Chun, and Ratnasamy}{Han
  et~al\mbox{.}}{2012}]%
        {han2012megapipe}
\bibfield{author}{\bibinfo{person}{Sangjin Han}, \bibinfo{person}{Scott
  Marshall}, \bibinfo{person}{Byung-Gon Chun}, {and} \bibinfo{person}{Sylvia
  Ratnasamy}.} \bibinfo{year}{2012}\natexlab{}.
\newblock \showarticletitle{MegaPipe: A New Programming Interface for Scalable
  Network I/O}. In \bibinfo{booktitle}{\emph{Proceedings of the 10th USENIX
  Conference on Operating Systems Design and Implementation}}.
\newblock


\bibitem[\protect\citeauthoryear{Hu, Song, and Li}{Hu et~al\mbox{.}}{2017}]%
        {hu_towards_2017}
\bibfield{author}{\bibinfo{person}{Yang Hu}, \bibinfo{person}{Mingcong Song},
  {and} \bibinfo{person}{Tao Li}.} \bibinfo{year}{2017}\natexlab{}.
\newblock \showarticletitle{Towards "{Full} {Containerization}" in
  {Containerized} {Network} {Function} {Virtualization}}. In
  \bibinfo{booktitle}{\emph{Proceedings of the {Twenty}-{Second}
  {International} {Conference} on {Architectural} {Support} for {Programming}
  {Languages} and {Operating} {Systems}}} \emph{(\bibinfo{series}{{ASPLOS}
  '17})}. \bibinfo{publisher}{ACM}, \bibinfo{address}{New York, NY, USA},
  \bibinfo{pages}{467--481}.
\newblock
\showISBNx{978-1-4503-4465-4}
\urldef\tempurl%
\url{https://doi.org/10.1145/3037697.3037713}
\showDOI{\tempurl}


\bibitem[\protect\citeauthoryear{intel}{intel}{[n. d.]}]%
        {noauthor_82599-10-gbe-controller-brief.pdf_nodate}
intel \bibinfo{year}{[n. d.]}\natexlab{}.
\newblock \bibinfo{title}{Intel 82599 10 gbe controller brief}.
\newblock
\newblock
\newblock
\shownote{https://www.intel.com/content/dam/www/public/us/en/documents/product-briefs/82599-10-gbe-controller-brief.pdf.}


\bibitem[\protect\citeauthoryear{Jeong, Wood, Jamshed, Jeong, Ihm, Han, and
  Park}{Jeong et~al\mbox{.}}{2014}]%
        {jeong_mtcp:_2014}
\bibfield{author}{\bibinfo{person}{EunYoung Jeong}, \bibinfo{person}{Shinae
  Wood}, \bibinfo{person}{Muhammad Jamshed}, \bibinfo{person}{Haewon Jeong},
  \bibinfo{person}{Sunghwan Ihm}, \bibinfo{person}{Dongsu Han}, {and}
  \bibinfo{person}{KyoungSoo Park}.} \bibinfo{year}{2014}\natexlab{}.
\newblock \showarticletitle{{mTCP}: a {Highly} {Scalable} {User}-level {TCP}
  {Stack} for {Multicore} {Systems}}. In \bibinfo{booktitle}{\emph{Proceedings
  of the 11th {USENIX} {Symposium} on {Networked} {Systems} {Design} and
  {Implementation}}} \emph{(\bibinfo{series}{{NSDI} 14})}.
  \bibinfo{publisher}{USENIX}, \bibinfo{address}{Seattle, WA},
  \bibinfo{pages}{489--502}.
\newblock
\showISBNx{978-1-931971-09-6}
\urldef\tempurl%
\url{https://www.usenix.org/conference/nsdi14/technical-sessions/presentation/jeong}
\showURL{%
\tempurl}


\bibitem[\protect\citeauthoryear{Kohler, Morris, Chen, Jannotti, and
  Kaashoek}{Kohler et~al\mbox{.}}{2000}]%
        {kohler00click}
\bibfield{author}{\bibinfo{person}{Eddie Kohler}, \bibinfo{person}{Robert
  Morris}, \bibinfo{person}{Benjie Chen}, \bibinfo{person}{John Jannotti},
  {and} \bibinfo{person}{M.~Frans Kaashoek}.} \bibinfo{year}{2000}\natexlab{}.
\newblock \showarticletitle{The Click modular router}.
\newblock \bibinfo{journal}{\emph{{ACM} Transactions on Computer Systems}}
  \bibinfo{volume}{18}, \bibinfo{number}{3} (\bibinfo{date}{August}
  \bibinfo{year}{2000}), \bibinfo{pages}{263--297}.
\newblock


\bibitem[\protect\citeauthoryear{Lagar-Cavilla, Whitney, Scannell, Patchin,
  Rumble, de~Lara, Brudno, and Satyanarayanan}{Lagar-Cavilla
  et~al\mbox{.}}{2009}]%
        {lagar_cavilla09snowflock}
\bibfield{author}{\bibinfo{person}{Horacio~Andr{\'e}s Lagar-Cavilla},
  \bibinfo{person}{Joseph~Andrew Whitney}, \bibinfo{person}{Adin~Matthew
  Scannell}, \bibinfo{person}{Philip Patchin}, \bibinfo{person}{Stephen~M.
  Rumble}, \bibinfo{person}{Eyal de Lara}, \bibinfo{person}{Michael Brudno},
  {and} \bibinfo{person}{Mahadev Satyanarayanan}.}
  \bibinfo{year}{2009}\natexlab{}.
\newblock \showarticletitle{SnowFlock: Rapid Virtual Machine Cloning for Cloud
  Computing}. In \bibinfo{booktitle}{\emph{Proceedings of the 4th ACM European
  Conference on Computer Systems (Eurosys)}}.
\newblock


\bibitem[\protect\citeauthoryear{Liedtke}{Liedtke}{1995}]%
        {liedtke95l4}
\bibfield{author}{\bibinfo{person}{J. Liedtke}.}
  \bibinfo{year}{1995}\natexlab{}.
\newblock \showarticletitle{On Micro-Kernel Construction}. In
  \bibinfo{booktitle}{\emph{Proceedings of the 15th {ACM} Symposium on
  Operating System Principles (SOSP'95), Copper Mountain Resort, Colorado, USA,
  December 3-6}}.
\newblock


\bibitem[\protect\citeauthoryear{Litton, Vahldiek-Oberwagner, Elnikety, Garg,
  Bhattacharjee, and Druschel}{Litton et~al\mbox{.}}{2016}]%
        {litton16lwcontexts}
\bibfield{author}{\bibinfo{person}{James Litton}, \bibinfo{person}{Anjo
  Vahldiek-Oberwagner}, \bibinfo{person}{Eslam Elnikety},
  \bibinfo{person}{Deepak Garg}, \bibinfo{person}{Bobby Bhattacharjee}, {and}
  \bibinfo{person}{Peter Druschel}.} \bibinfo{year}{2016}\natexlab{}.
\newblock \showarticletitle{Light-weight Contexts: An OS Abstraction for Safety
  and Performance}. In \bibinfo{booktitle}{\emph{Proceedings of the 12th USENIX
  Conference on Operating Systems Design and Implementation (OSDI)}}.
\newblock


\bibitem[\protect\citeauthoryear{Madhavapeddy, Leonard, Skjegstad, Gazagnaire,
  Sheets, Scott, Mortier, Chaudhry, Singh, Ludlam, Crowcroft, and
  Leslie}{Madhavapeddy et~al\mbox{.}}{2015}]%
        {madhavapeddy15jitsu}
\bibfield{author}{\bibinfo{person}{Anil Madhavapeddy}, \bibinfo{person}{Thomas
  Leonard}, \bibinfo{person}{Magnus Skjegstad}, \bibinfo{person}{Thomas
  Gazagnaire}, \bibinfo{person}{David Sheets}, \bibinfo{person}{Dave Scott},
  \bibinfo{person}{Richard Mortier}, \bibinfo{person}{Amir Chaudhry},
  \bibinfo{person}{Balraj Singh}, \bibinfo{person}{Jon Ludlam},
  \bibinfo{person}{Jon Crowcroft}, {and} \bibinfo{person}{Ian Leslie}.}
  \bibinfo{year}{2015}\natexlab{}.
\newblock \showarticletitle{Jitsu: {Just}-in-time {Summoning} of {Unikernels}}.
  In \bibinfo{booktitle}{\emph{Proceedings of the 12th {USENIX} {Conference} on
  {Networked} {Systems} {Design} and {Implementation}}}
  \emph{(\bibinfo{series}{{NSDI}'15})}. \bibinfo{publisher}{USENIX
  Association}, \bibinfo{address}{Berkeley, CA, USA},
  \bibinfo{pages}{559--573}.
\newblock
\showISBNx{978-1-931971-21-8}
\urldef\tempurl%
\url{http://dl.acm.org/citation.cfm?id=2789770.2789809}
\showURL{%
\tempurl}


\bibitem[\protect\citeauthoryear{Madhavapeddy, Mortier, Rotsos, Scott, Singh,
  Gazagnaire, Smith, Hand, and Crowcroft}{Madhavapeddy et~al\mbox{.}}{2013}]%
        {madhavapeddy13unikernels}
\bibfield{author}{\bibinfo{person}{Anil Madhavapeddy}, \bibinfo{person}{Richard
  Mortier}, \bibinfo{person}{Charalampos Rotsos}, \bibinfo{person}{David
  Scott}, \bibinfo{person}{Balraj Singh}, \bibinfo{person}{Thomas Gazagnaire},
  \bibinfo{person}{Steven Smith}, \bibinfo{person}{Steven Hand}, {and}
  \bibinfo{person}{Jon Crowcroft}.} \bibinfo{year}{2013}\natexlab{}.
\newblock \showarticletitle{Unikernels: {Library} {Operating} {Systems} for the
  {Cloud}}. In \bibinfo{booktitle}{\emph{Proceedings of the {Eighteenth}
  {International} {Conference} on {Architectural} {Support} for {Programming}
  {Languages} and {Operating} {Systems}}} \emph{(\bibinfo{series}{{ASPLOS}
  '13})}. \bibinfo{publisher}{ACM}, \bibinfo{address}{New York, NY, USA},
  \bibinfo{pages}{461--472}.
\newblock
\showISBNx{978-1-4503-1870-9}
\urldef\tempurl%
\url{https://doi.org/10.1145/2451116.2451167}
\showDOI{\tempurl}


\bibitem[\protect\citeauthoryear{Manco, Lupu, Schmidt, Mendes, Kuenzer, Sati,
  Yasukata, Raiciu, and Huici}{Manco et~al\mbox{.}}{2017}]%
        {manco17lightvm}
\bibfield{author}{\bibinfo{person}{Filipe Manco}, \bibinfo{person}{Costin
  Lupu}, \bibinfo{person}{Florian Schmidt}, \bibinfo{person}{Jose Mendes},
  \bibinfo{person}{Simon Kuenzer}, \bibinfo{person}{Sumit Sati},
  \bibinfo{person}{Kenichi Yasukata}, \bibinfo{person}{Costin Raiciu}, {and}
  \bibinfo{person}{Felipe Huici}.} \bibinfo{year}{2017}\natexlab{}.
\newblock \showarticletitle{My VM is Lighter (and Safer) Than Your Container}.
  In \bibinfo{booktitle}{\emph{Proceedings of the 26th Symposium on Operating
  Systems Principles (SOSP)}}.
\newblock


\bibitem[\protect\citeauthoryear{Martins, Ahmed, Raiciu, Olteanu, Honda,
  Bifulco, and Huici}{Martins et~al\mbox{.}}{2014}]%
        {martins14clickos}
\bibfield{author}{\bibinfo{person}{Joao Martins}, \bibinfo{person}{Mohamed
  Ahmed}, \bibinfo{person}{Costin Raiciu}, \bibinfo{person}{Vladimir Olteanu},
  \bibinfo{person}{Michio Honda}, \bibinfo{person}{Roberto Bifulco}, {and}
  \bibinfo{person}{Felipe Huici}.} \bibinfo{year}{2014}\natexlab{}.
\newblock \showarticletitle{ClickOS and the Art of Network Function
  Virtualization}. In \bibinfo{booktitle}{\emph{Proceedings of the 11th USENIX
  Conference on Networked Systems Design and Implementation (NSDI)}}.
\newblock


\bibitem[\protect\citeauthoryear{Miller, Yee, and Shapiro}{Miller
  et~al\mbox{.}}{2003}]%
        {miller03cap_myths}
\bibfield{author}{\bibinfo{person}{Mark~S. Miller}, \bibinfo{person}{Ka-Ping
  Yee}, {and} \bibinfo{person}{Jonathan Shapiro}.}
  \bibinfo{year}{2003}\natexlab{}.
\newblock \bibinfo{booktitle}{\emph{Capability myths demolished}}.
\newblock \bibinfo{type}{{T}echnical {R}eport} SRL2003-02.
  \bibinfo{institution}{Johns Hopkins University Systems Research Laboratory},
  \bibinfo{address}{Mountain View CA ({USA})}.
\newblock
\urldef\tempurl%
\url{http://www.erights.org/elib/capability/duals/}
\showURL{%
\tempurl}


\bibitem[\protect\citeauthoryear{Nishtala, Fugal, Grimm, Kwiatkowski, Lee, Li,
  McElroy, Paleczny, Peek, Saab, Stafford, Tung, and Venkataramani}{Nishtala
  et~al\mbox{.}}{2013}]%
        {nishtala_scaling_2013}
\bibfield{author}{\bibinfo{person}{Rajesh Nishtala}, \bibinfo{person}{Hans
  Fugal}, \bibinfo{person}{Steven Grimm}, \bibinfo{person}{Marc Kwiatkowski},
  \bibinfo{person}{Herman Lee}, \bibinfo{person}{Harry~C. Li},
  \bibinfo{person}{Ryan McElroy}, \bibinfo{person}{Mike Paleczny},
  \bibinfo{person}{Daniel Peek}, \bibinfo{person}{Paul Saab},
  \bibinfo{person}{David Stafford}, \bibinfo{person}{Tony Tung}, {and}
  \bibinfo{person}{Venkateshwaran Venkataramani}.}
  \bibinfo{year}{2013}\natexlab{}.
\newblock \showarticletitle{Scaling {Memcache} at {Facebook}}. In
  \bibinfo{booktitle}{\emph{Presented as part of the 10th {USENIX} {Symposium}
  on {Networked} {Systems} {Design} and {Implementation} ({NSDI} 13)}}.
  \bibinfo{publisher}{USENIX}, \bibinfo{address}{Lombard, IL},
  \bibinfo{pages}{385--398}.
\newblock
\showISBNx{978-1-931971-00-3}
\urldef\tempurl%
\url{https://www.usenix.org/conference/nsdi13/technical-sessions/presentation/nishtala}
\showURL{%
\tempurl}


\bibitem[\protect\citeauthoryear{Nitu, Olivier, Tchana, Chiba, Barbalace,
  Hagimont, and Ravindran}{Nitu et~al\mbox{.}}{2017}]%
        {nitu_swift_2017}
\bibfield{author}{\bibinfo{person}{Vlad Nitu}, \bibinfo{person}{Pierre
  Olivier}, \bibinfo{person}{Alain Tchana}, \bibinfo{person}{Daniel Chiba},
  \bibinfo{person}{Antonio Barbalace}, \bibinfo{person}{Daniel Hagimont}, {and}
  \bibinfo{person}{Binoy Ravindran}.} \bibinfo{year}{2017}\natexlab{}.
\newblock \showarticletitle{Swift {Birth} and {Quick} {Death}: {Enabling}
  {Fast} {Parallel} {Guest} {Boot} and {Destruction} in the {Xen}
  {Hypervisor}}. In \bibinfo{booktitle}{\emph{Proceedings of the 13th {ACM}
  {SIGPLAN}/{SIGOPS} {International} {Conference} on {Virtual} {Execution}
  {Environments}}} \emph{(\bibinfo{series}{{VEE} '17})}.
  \bibinfo{publisher}{ACM}, \bibinfo{address}{New York, NY, USA},
  \bibinfo{pages}{1--14}.
\newblock
\showISBNx{978-1-4503-4948-2}
\urldef\tempurl%
\url{https://doi.org/10.1145/3050748.3050758}
\showDOI{\tempurl}


\bibitem[\protect\citeauthoryear{Oakes, Yang, Zhou, Houck, Harter,
  Arpaci-Dusseau, and Arpaci-Dusseau}{Oakes et~al\mbox{.}}{2018}]%
        {oakes18sock}
\bibfield{author}{\bibinfo{person}{Edward Oakes}, \bibinfo{person}{Leon Yang},
  \bibinfo{person}{Dennis Zhou}, \bibinfo{person}{Kevin Houck},
  \bibinfo{person}{Tyler Harter}, \bibinfo{person}{Andrea Arpaci-Dusseau},
  {and} \bibinfo{person}{Remzi Arpaci-Dusseau}.}
  \bibinfo{year}{2018}\natexlab{}.
\newblock \showarticletitle{{SOCK}: Rapid Task Provisioning with
  Serverless-Optimized Containers}. In \bibinfo{booktitle}{\emph{2018 {USENIX}
  Annual Technical Conference ({USENIX} {ATC} 18)}}.
\newblock


\bibitem[\protect\citeauthoryear{Palkar, Lan, Han, Jang, Panda, Ratnasamy,
  Rizzo, and Shenker}{Palkar et~al\mbox{.}}{2015}]%
        {palkar15e2}
\bibfield{author}{\bibinfo{person}{Shoumik Palkar}, \bibinfo{person}{Chang
  Lan}, \bibinfo{person}{Sangjin Han}, \bibinfo{person}{Keon Jang},
  \bibinfo{person}{Aurojit Panda}, \bibinfo{person}{Sylvia Ratnasamy},
  \bibinfo{person}{Luigi Rizzo}, {and} \bibinfo{person}{Scott Shenker}.}
  \bibinfo{year}{2015}\natexlab{}.
\newblock \showarticletitle{E2: A Framework for NFV Applications}. In
  \bibinfo{booktitle}{\emph{Proceedings of the 25th Symposium on Operating
  Systems Principles (SOSP)}}.
\newblock


\bibitem[\protect\citeauthoryear{Panda, Han, Jang, Walls, Ratnasamy, and
  Shenker}{Panda et~al\mbox{.}}{2016}]%
        {panda16netbricks}
\bibfield{author}{\bibinfo{person}{Aurojit Panda}, \bibinfo{person}{Sangjin
  Han}, \bibinfo{person}{Keon Jang}, \bibinfo{person}{Melvin Walls},
  \bibinfo{person}{Sylvia Ratnasamy}, {and} \bibinfo{person}{Scott Shenker}.}
  \bibinfo{year}{2016}\natexlab{}.
\newblock \showarticletitle{NetBricks: Taking the V out of NFV}. In
  \bibinfo{booktitle}{\emph{Proceedings of the 12th USENIX Conference on
  Operating Systems Design and Implementation (OSDI)}}.
\newblock


\bibitem[\protect\citeauthoryear{Parmer and West}{Parmer and West}{2008}]%
        {parmer08composite_sched}
\bibfield{author}{\bibinfo{person}{Gabriel Parmer} {and}
  \bibinfo{person}{Richard West}.} \bibinfo{year}{2008}\natexlab{}.
\newblock \showarticletitle{Predictable Interrupt Management and Scheduling in
  the {C}omposite Component-based System}. In
  \bibinfo{booktitle}{\emph{Proceedings of the 29th {IEEE} Real-Time Systems
  Symposium (RTSS'08), Barcelona, Spain, November 30 - December 3}}.
\newblock


\bibitem[\protect\citeauthoryear{Parmer and West}{Parmer and West}{2011}]%
        {parmer11hires}
\bibfield{author}{\bibinfo{person}{Gabriel Parmer} {and}
  \bibinfo{person}{Richard West}.} \bibinfo{year}{2011}\natexlab{}.
\newblock \showarticletitle{Hi{R}es: A System for Predictable Hierarchical
  Resource Management}. In \bibinfo{booktitle}{\emph{Proceedings of the 17th
  IEEE Real-Time and Embedded Technology and Applications Symposium (RTAS)}}.
\newblock


\bibitem[\protect\citeauthoryear{Peter, Li, Zhang, Ports, Woos, Krishnamurthy,
  Anderson, and Roscoe}{Peter et~al\mbox{.}}{2015}]%
        {peter15arrakis}
\bibfield{author}{\bibinfo{person}{Simon Peter}, \bibinfo{person}{Jialin Li},
  \bibinfo{person}{Irene Zhang}, \bibinfo{person}{Dan R.~K. Ports},
  \bibinfo{person}{Doug Woos}, \bibinfo{person}{Arvind Krishnamurthy},
  \bibinfo{person}{Thomas Anderson}, {and} \bibinfo{person}{Timothy Roscoe}.}
  \bibinfo{year}{2015}\natexlab{}.
\newblock \showarticletitle{Arrakis: The Operating System Is the Control
  Plane}.
\newblock \bibinfo{journal}{\emph{ACM Trans. Comput. Syst.}}
  \bibinfo{volume}{33}, \bibinfo{number}{4} (\bibinfo{date}{Nov.}
  \bibinfo{year}{2015}).
\newblock


\bibitem[\protect\citeauthoryear{Peterson}{Peterson}{2015}]%
        {peterson_cord:_2015}
\bibfield{author}{\bibinfo{person}{Larry Peterson}.}
  \bibinfo{year}{2015}\natexlab{}.
\newblock \showarticletitle{Cord: {Central} office re-architected as a
  datacenter}.
\newblock \bibinfo{journal}{\emph{Open Networking Lab white paper}}
  (\bibinfo{year}{2015}).
\newblock


\bibitem[\protect\citeauthoryear{Price and Tucker}{Price and Tucker}{2004}]%
        {price04zones}
\bibfield{author}{\bibinfo{person}{Daniel Price} {and} \bibinfo{person}{Andrew
  Tucker}.} \bibinfo{year}{2004}\natexlab{}.
\newblock \showarticletitle{Solaris Zones: Operating System Support for
  Consolidating Commercial Workloads}. In \bibinfo{booktitle}{\emph{Proceedings
  of the 18th USENIX Conference on System Administration (LISA)}}.
\newblock


\bibitem[\protect\citeauthoryear{Rizzo}{Rizzo}{2012}]%
        {rizzo12netmap}
\bibfield{author}{\bibinfo{person}{Luigi Rizzo}.}
  \bibinfo{year}{2012}\natexlab{}.
\newblock \showarticletitle{Netmap: A Novel Framework for Fast Packet I/O}. In
  \bibinfo{booktitle}{\emph{Proceedings of the 2012 USENIX Conference on Annual
  Technical Conference (USENIX ATC)}}.
\newblock


\bibitem[\protect\citeauthoryear{Saltzer and Schroeder}{Saltzer and
  Schroeder}{1975}]%
        {saltzer75info_protection}
\bibfield{author}{\bibinfo{person}{J. Saltzer} {and} \bibinfo{person}{M.
  Schroeder}.} \bibinfo{year}{1975}\natexlab{}.
\newblock \showarticletitle{The protection of information in computer systems}.
\newblock \bibinfo{journal}{\emph{in Proceedings of the {IEEE}}}
  \bibinfo{volume}{9}, \bibinfo{number}{63} (\bibinfo{year}{1975}).
\newblock


\bibitem[\protect\citeauthoryear{Shapiro, Smith, and Farber}{Shapiro
  et~al\mbox{.}}{1999}]%
        {shapiro99eros}
\bibfield{author}{\bibinfo{person}{Jonathan~S. Shapiro},
  \bibinfo{person}{Jonathan~M. Smith}, {and} \bibinfo{person}{David~J.
  Farber}.} \bibinfo{year}{1999}\natexlab{}.
\newblock \showarticletitle{{EROS}: a fast capability system}. In
  \bibinfo{booktitle}{\emph{Proceedings of the 17th {ACM} Symposium on
  Operating System Principles (SOSP'99), Kiawah Island Resort, South Carolina,
  USA, December 12-15}}.
\newblock


\bibitem[\protect\citeauthoryear{Sharma, Kaufmann, Anderson, Kim,
  Krishnamurthy, Nelson, and Peter}{Sharma et~al\mbox{.}}{2017}]%
        {sharma17flexnic}
\bibfield{author}{\bibinfo{person}{Naveen~Kr. Sharma}, \bibinfo{person}{Antoine
  Kaufmann}, \bibinfo{person}{Thomas Anderson}, \bibinfo{person}{Changhoon
  Kim}, \bibinfo{person}{Arvind Krishnamurthy}, \bibinfo{person}{Jacob Nelson},
  {and} \bibinfo{person}{Simon Peter}.} \bibinfo{year}{2017}\natexlab{}.
\newblock \showarticletitle{Evaluating the Power of Flexible Packet Processing
  for Network Resource Allocation}. In \bibinfo{booktitle}{\emph{Proceedings of
  the 14th USENIX Conference on Networked Systems Design and Implementation
  (NSDI)}}.
\newblock


\bibitem[\protect\citeauthoryear{Taleb, Samdanis, Mada, Flinck, Dutta, and
  Sabella}{Taleb et~al\mbox{.}}{2017}]%
        {taleb_multi-access_2017}
\bibfield{author}{\bibinfo{person}{T. Taleb}, \bibinfo{person}{K. Samdanis},
  \bibinfo{person}{B. Mada}, \bibinfo{person}{H. Flinck}, \bibinfo{person}{S.
  Dutta}, {and} \bibinfo{person}{D. Sabella}.} \bibinfo{year}{2017}\natexlab{}.
\newblock \showarticletitle{On {Multi}-{Access} {Edge} {Computing}: {A}
  {Survey} of the {Emerging} 5G {Network} {Edge} {Cloud} {Architecture} and
  {Orchestration}}.
\newblock \bibinfo{journal}{\emph{IEEE Communications Surveys Tutorials}}
  \bibinfo{volume}{19}, \bibinfo{number}{3} (\bibinfo{year}{2017}),
  \bibinfo{pages}{1657--1681}.
\newblock
\urldef\tempurl%
\url{https://doi.org/10.1109/COMST.2017.2705720}
\showDOI{\tempurl}


\bibitem[\protect\citeauthoryear{Tennenhouse}{Tennenhouse}{1989}]%
        {tennenhouse89layers_harmful}
\bibfield{author}{\bibinfo{person}{David Tennenhouse}.}
  \bibinfo{year}{1989}\natexlab{}.
\newblock \showarticletitle{Layered Multiplexing Considered Harmful}. In
  \bibinfo{booktitle}{\emph{Protocols for High-Speed Networks}}.
  \bibinfo{address}{North Holland, Amsterdam}, \bibinfo{pages}{143--148}.
\newblock


\bibitem[\protect\citeauthoryear{Thalheim, Bhatotia, Fonseca, and
  Kasikci}{Thalheim et~al\mbox{.}}{2018}]%
        {thalheim18cntr}
\bibfield{author}{\bibinfo{person}{J{\"o}rg Thalheim}, \bibinfo{person}{Pramod
  Bhatotia}, \bibinfo{person}{Pedro Fonseca}, {and} \bibinfo{person}{Baris
  Kasikci}.} \bibinfo{year}{2018}\natexlab{}.
\newblock \showarticletitle{Cntr: Lightweight {OS} Containers}. In
  \bibinfo{booktitle}{\emph{2018 {USENIX} Annual Technical Conference ({USENIX}
  {ATC} 18)}}.
\newblock


\bibitem[\protect\citeauthoryear{Vilanova, Jord\`{a}, Navarro, Etsion, and
  Valero}{Vilanova et~al\mbox{.}}{2017}]%
        {vilanova17dipc}
\bibfield{author}{\bibinfo{person}{Llu\'{\i}s Vilanova}, \bibinfo{person}{Marc
  Jord\`{a}}, \bibinfo{person}{Nacho Navarro}, \bibinfo{person}{Yoav Etsion},
  {and} \bibinfo{person}{Mateo Valero}.} \bibinfo{year}{2017}\natexlab{}.
\newblock \showarticletitle{Direct Inter-Process Communication (dIPC):
  Repurposing the CODOMs Architecture to Accelerate IPC}. In
  \bibinfo{booktitle}{\emph{Proceedings of the Twelfth European Conference on
  Computer Systems (Eurosys)}}.
\newblock


\bibitem[\protect\citeauthoryear{von Eicken, Basu, Buch, and Vogels}{von Eicken
  et~al\mbox{.}}{1995}]%
        {eicken95unet}
\bibfield{author}{\bibinfo{person}{Thorsten von Eicken},
  \bibinfo{person}{Anindya Basu}, \bibinfo{person}{Vineet Buch}, {and}
  \bibinfo{person}{Werner Vogels}.} \bibinfo{year}{1995}\natexlab{}.
\newblock \showarticletitle{{U}-{N}et: A User-Level Network Interface for
  Parallel and Distributed Computing}. In \bibinfo{booktitle}{\emph{Proceedings
  of the 14th ACM Symposium on Operating Systems Principles}}. ACM,
  \bibinfo{pages}{40--53}.
\newblock


\bibitem[\protect\citeauthoryear{Wang, Ren, Scaperoth, and Parmer}{Wang
  et~al\mbox{.}}{2015}]%
        {wang15speck}
\bibfield{author}{\bibinfo{person}{Qi Wang}, \bibinfo{person}{Yuxin Ren},
  \bibinfo{person}{Matt Scaperoth}, {and} \bibinfo{person}{Gabriel Parmer}.}
  \bibinfo{year}{2015}\natexlab{}.
\newblock \showarticletitle{Speck: A Kernel for Scalable Predictability}. In
  \bibinfo{booktitle}{\emph{Proceedings of the 21st {IEEE} Real-Time and
  Embedded Technology and Applications Symposium (RTAS'15), Seattle, WA, USA,
  April 13-16}}.
\newblock


\bibitem[\protect\citeauthoryear{Whitaker, Shaw, and Gribble}{Whitaker
  et~al\mbox{.}}{2002}]%
        {whitaker02denali}
\bibfield{author}{\bibinfo{person}{A. Whitaker}, \bibinfo{person}{M. Shaw},
  {and} \bibinfo{person}{S. Gribble}.} \bibinfo{year}{2002}\natexlab{}.
\newblock \bibinfo{title}{Denali: Lightweight virtual machines for distributed
  and networked applications}.
\newblock
\newblock
\urldef\tempurl%
\url{citeseer.ist.psu.edu/whitaker02denali.html}
\showURL{%
\tempurl}


\bibitem[\protect\citeauthoryear{Zhang, Liu, Zhang, Shah, Lopreiato, Todeschi,
  Ramakrishnan, and Wood}{Zhang et~al\mbox{.}}{2016}]%
        {zhang_opennetvm:_2016}
\bibfield{author}{\bibinfo{person}{Wei Zhang}, \bibinfo{person}{Guyue Liu},
  \bibinfo{person}{Wenhui Zhang}, \bibinfo{person}{Neel Shah},
  \bibinfo{person}{Phillip Lopreiato}, \bibinfo{person}{Gregoire Todeschi},
  \bibinfo{person}{K.K. Ramakrishnan}, {and} \bibinfo{person}{Timothy Wood}.}
  \bibinfo{year}{2016}\natexlab{}.
\newblock \showarticletitle{{OpenNetVM}: {A} {Platform} for {High}
  {Performance} {Network} {Service} {Chains}}. In
  \bibinfo{booktitle}{\emph{Proceedings of the 2016 {ACM} {SIGCOMM} {Workshop}
  on {Hot} {Topics} in {Middleboxes} and {Network} {Function}
  {Virtualization}}}. \bibinfo{publisher}{ACM}.
\newblock
\urldef\tempurl%
\url{http://faculty.cs.gwu.edu/ timwood/papers/16-HotMiddlebox-onvm.pdf}
\showURL{%
\tempurl}


\end{thebibliography}
